\documentstyle{mn2e}
\input psfig

  \makeatletter
  \ifx\CUP@mtlplain@loaded\undefined  
  \makeatother  
    
  \else
    
  \fi

\loadboldmathitalic 
\loadboldgreek      
  
\makeatletter
\ifx\CUP@mtlplain@loaded\undefined  
  \newfont\bit{cmbxti10 at 9pt}
\else
  \newfont\bit{mtbxti10 at 9pt}
\fi
\makeatother

\def\LaTeX{L\kern-.36em\raise.3ex\hbox{a}\kern-.15em
    T\kern-.1667em\lower.7ex\hbox{E}\kern-.125emX}

\newcommand{\gsim}{\mathrel{\hbox{\rlap{\lower.55ex \hbox {$\sim$}}
                   \kern-.3em \raise.4ex \hbox{$>$}}}}
\newcommand{\lsim}{\mathrel{\hbox{\rlap{\lower.55ex \hbox {$\sim$}}
                   \kern-.3em \raise.4ex \hbox{$<$}}}}

\title[The properties of stars and brown dwarfs]{The formation of a star cluster: predicting the properties of stars and brown dwarfs}

\author[M. R. Bate et~al.]
  {Matthew R. Bate,$^{1,2}$\thanks{E-mail: mbate@astro.ex.ac.uk}
  Ian A. Bonnell,$^3$
  and Volker Bromm.$^{2,4}$\\
  $^1$School of Physics, University of Exeter, Stocker Road,
    Exeter EX4 4QL \\
  $^2$Institute of Astronomy, University of Cambridge, Madingley Road,
    Cambridge CB3 0HA \\
  $^3$School of Physics and Astronomy, University of St Andrews, North Haugh, St Andrews, Fife, KY16 9SS \\
  $^4$Harvard-Smithsonian Center for Astrophysics, 60 Garden Street, Cambridge,
MA 02138, U.S.A.
}

\date{Accepted for publication in MNRAS}

\begin{document}

\maketitle

\begin{abstract}
  We present results from the largest numerical simulation of star 
  formation to resolve the fragmentation process down to the 
  opacity limit.  The simulation follows the collapse and 
  fragmentation of a large-scale turbulent molecular cloud to form
  a stellar cluster and, simultaneously, the formation of circumstellar 
  discs and binary stars.  This large range of scales
  enables us to predict a wide variety of stellar properties for 
  comparison with observations.

  The calculation clearly demonstrates that star formation is a 
  highly-dynamic and chaotic process.  Star formation occurs in 
  localised bursts within the cloud via the fragmentation both
  of dense molecular cloud cores and of massive circumstellar discs.
  Star-disc encounters form binaries and truncate discs.
  Stellar encounters disrupt bound multiple systems.
  We find that the observed 
  statistical properties of stars are a natural consequence of 
  star formation in such a dynamical environment.  
  The cloud produces roughly equal numbers of 
  stars and brown dwarfs, with masses down to the
  opacity limit for fragmentation ($\approx 5$ Jupiter masses).
  The initial mass function is consistent with a Salpeter slope 
  ($\Gamma=-1.35$) above 0.5 M$_\odot$, a roughly flat distribution 
  ($\Gamma=0$) in the range $0.006-0.5$ M$_\odot$, and a sharp 
  cutoff below $\approx 0.005$ M$_\odot$.
  This is consistent with recent observational surveys.
  The brown dwarfs form by the dynamical ejection of low-mass fragments
  from dynamically unstable multiple systems before the fragments 
  have been able to accrete to stellar masses.  
  Close binary systems (with separations $\lsim 10$ AU) are not formed 
  by fragmentation in situ.  Rather, they are produced by hardening of
  initially wider multiple systems through a combination of 
  dynamical encounters, gas accretion, and/or the interaction with
  circumbinary and circumtriple discs.  
  Finally, we find that the majority of circumstellar discs have 
  radii less than 20 AU due to truncation in dynamical encounters.
  This is consistent with observations of 
  the Orion Trapezium Cluster and implies that most stars and 
  brown dwarfs do not form large planetary systems.
\end{abstract}

\begin{keywords}
  accretion, accretion discs -- binaries: general -- brown dwarfs -- hydrodynamics -- stars: formation -- stars: mass function.
\end{keywords}

\section{Introduction}

The collapse and fragmentation of molecular cloud cores to form 
bound multiple stellar systems has been the subject of
many numerical studies (e.g.\ Boss \& Bodenheimer 1979;
Boss 1986; Bonnell et al.\ 1991; Nelson \& Papaloizou 1993; Bonnell 1994; 
Burkert \& Bodenheimer 1993; Bate, Bonnell \& Price 1995; 
Truelove et al.\ 1998).
These calculations have resulted in the adoption of 
fragmentation as the favoured mechanism for the 
formation of binary and multiple stars, since it can produce 
a wide range of binary properties through simple 
variations of the pre-collapse initial conditions.

However, while individual binary systems can be reproduced 
by such fragmentation calculations, it is extremely difficult 
to use these calculations to predict the statistical properties 
of the stellar systems that should result from the fragmentation model.
Quantities that we may wish to determine include the initial
mass function (IMF), the relative frequencies of single, binary 
and multiple stars, the properties of multiple stars, the properties
of circumstellar discs, and the efficiency of star formation.
In order to predict these statistical properties, we need
to produce a large sample of stars.  There are two possibilities.

We could perform many calculations of isolated cloud cores
using a representative sample of initial conditions.  However, this
has two disadvantages.  First, the conditions in molecular clouds
are not sufficiently well understood to be able to select a representative
sample of cloud cores for the initial conditions.  Second, the 
production of isolated stellar systems neglects interactions 
between systems that may be important in determining stellar
properties, especially in young star clusters.  Examples of such
interactions include binary formation via star-disc capture
(Larson 1990; Clarke \& Pringle 1991a,b; Heller 1995;
McDonald \& Clarke 1995; Hall, Clarke \& Pringle 1996), 
truncation of protostellar discs (Heller 1995; Hall 1997), 
and competitive accretion leading to a range of stellar masses
(Larson 1978; Zinnecker 1982; Bonnell et al. 1997, 2001a,b).

The second possibility is to perform a calculation
of the collapse and fragmentation of a large-scale molecular
cloud to form many stars simultaneously.  This is the approach 
taken in this paper.  Interactions between stars are automatically
allowed for.  We must still specify global initial conditions, but the
formation of individual cores within the cloud occurs
self-consistently; we do not have to select initial
conditions for each core arbitrarily.  The only disadvantage
is that such a calculation is extremely computationally intensive.

Several such global calculations have been performed in the past.
The earliest was that of Chapman et al.\ \shortcite{Chapmanetal1992}
who followed the collapse and fragmentation of a shock-compressed
layer of molecular gas between two colliding clouds.  The calculation
produced many single, binary, and multiple protostars, but there was 
no attempt to derive the statistical properties of these systems.
Klessen, Burkert \& Bate \shortcite{KleBurBat1998}
followed the collapse of a large-scale clumpy molecular cloud 
to form $\sim 60$ protostars.  They found that the mass 
function of the protostars could be fit by a lognormal 
mass function that has a similar width to the observed stellar 
initial mass function.  The protostellar masses were set 
by a combination of the initial density structure, competitive 
accretion, and dynamical interactions.  Further calculations 
in which the global initial conditions were varied confirmed the 
lognormal form of the mass function and showed that the mean mass of the 
protostars was similar to the mean initial Jeans mass 
in the cloud (Klessen \& Burkert 2000, 2001; Klessen 2001).
These calculations enabled us to identify some of the processes 
that may help to determine the initial mass function.  However,
they did not have the resolution to follow the collapsing molecular
gas all the way down to the opacity limit for fragmentation
(Low \& Lynden-Bell 1976; Rees 1976; Silk 1977a), 
or even to resolve the median separation of binary systems
of $\approx 30$ AU \cite{DuqMay1991}.  
Thus, we cannot determine the total number of stars that will 
form or the stellar initial mass function from these calculations, 
let alone the frequency of binary stars or the importance 
of star-disc encounters.

This paper presents results from the first calculation 
to follow the collapse and fragmentation of a large-scale 
turbulent molecular cloud to form a stellar cluster
while resolving beyond the opacity limit for fragmentation.
Thus, assuming that fragmentation does not occur at densities 
greater than those at which the opacity limit sets in (Section
\ref{opacitysec}), it resolves
all potential fragmentation, including that which produces 
binary systems.  This allows us to predict a wide variety of 
stellar properties.  Two papers that contain results from this 
calculation have already been published.  They concentrate 
on the formation mechanisms of brown dwarfs 
(Bate, Bonnell \& Bromm 2002a) and close binaries 
(Bate, Bonnell \& Bromm 2002b).  In this paper, we consider how the
dynamics of star formation determine the properties of stars and 
brown dwarfs, and we compare these properties with observations.

The outline of this paper is as follows.  In Section 2, we briefly
review the opacity limit for fragmentation.  The computational 
method and the initial conditions for our calculation are 
described in Section 3.  Section 4 discusses the evolution of the
cloud and the star formation that occurs during the calculation.  
The properties of the resulting stars and brown dwarfs are 
compared with observations of star-forming regions in Section 5.  
Finally, in section 6, we give our conclusions

\section{The opacity limit for fragmentation}
\label{opacitysec}

When a molecular cloud core begins to collapse under its own gravity
the gravitational potential energy that is released can easily 
be radiated away so that the collapsing gas is approximately 
isothermal (e.g.\ Larson 1969).  
Thus, the thermal pressure varies with density $\rho$ as
$p \propto \rho^\eta$ where the effective polytropic 
exponent $\eta \equiv {\rm d}\log[p]/{\rm d}\log[\rho] \approx 1$.
This allows the possibility of fragmentation because the Jeans 
mass decreases with increasing density if $\eta <4/3$.

The opacity limit for fragmentation occurs when the rate at which energy 
is released by the collapse exceeds the rate at which energy can be
radiated away (Rees 1976; Low \& Lynden-Bell 1976; 
Masunaga \& Inutsuka 1999).  The gas then heats up with
$\eta > 4/3$, the Jeans mass increases, and a Jeans-unstable
collapsing clump quickly becomes Jeans-stable so that 
a pressure-supported fragment forms.  The density at which this
occurs depends on the opacity (and the initial temperature) of the gas
(Masunaga \& Inutsuka 1999), hence the term `opacity limit for 
fragmentation'.  For standard molecular gas at an initial temperature
of 10 K the gas begins to heat significantly 
at a density of $\approx 10^{-13}~{\rm g~cm}^{-3}$ 
(Larson 1969; Masunaga \& Inutsuka 2000).

The pressure-supported fragment initially contains several
Jupiter-masses (M$_{\rm J}$) and has a radius of $\approx 5$ AU
\cite{Larson1969}.  Such a fragment is expected to be
embedded within a collapsing envelope.  Thus, its mass grows
with time and its central temperature increases.
When its central temperature reaches 2000 K, molecular hydrogen 
begins to dissociate.  This provides a way for the release 
of gravitational energy to be absorbed without significantly 
increasing the temperature of the gas.  Thus, a nearly 
isothermal second collapse occurs within the fragment that
ultimately results in the formation of a stellar core with 
radius $\approx 1 R_\odot$ \cite{Larson1969}.

Several studies have investigated the possibility of fragmentation 
during this second collapse (Boss 1989; Bonnell \& Bate 1994; 
Bate 1998, 2002).  Boss \shortcite{Boss1989} found that 
fragmentation was possible during this second collapse, but that 
the objects spiralled together and merged due to gravitational 
torques from non-axisymmetric structure.  
Bonnell \& Bate \shortcite{BonBat1994} found that fragmentation
to form close binaries and multiple systems could occur in a disc 
that forms around the stellar core.  However, both these studies
began with somewhat arbitrary initial conditions and modelled only the 
pressure-supported fragment.  Bate \shortcite{Bate1998} performed 
the first three-dimensional calculations to follow the entire collapse 
of a molecular cloud core through the formation of the 
pressure-supported fragment, the second collapse phase,
and the formation of the stellar core and its surrounding disc.
In these and subsequent calculations (Bate 2002), it was found 
that the second collapse did not result in sub-fragmentation 
due to the high degree of thermal pressure and 
angular momentum transport via gravitational torques 
from non-axisymmetric structure.

Therefore, it appears that the opacity limit for fragmentation is real
and that fragmentation cannot occur at densities exceeding 
$\approx 10^{-13}~{\rm g~cm}^{-3}$.  The implications are that
there should be a minimum `stellar' mass of $\sim 10$ M$_{\rm J}$
(Low \& Lynden-Bell 1976; Silk 1977a, Boss 1988) 
and that protobinaries should have a minimum initial 
separation of $\approx 10$ AU due to the sizes of the 
pressure-supported fragments.  The exact value of the 
minimum mass is uncertain with theoretical values in the
literature ranging from $1-10$ M$_{\rm J}$ (Low \& Lynden-Bell 1976; 
Silk 1977a; Boss 1988; Masunaga \& Inutsuka 1999; Boss 2001).  
Surveys of young star clusters are beginning to probe 
masses down to this theoretical minimum
mass (Zapatero Osorio et al.\ 1999; Lucas \& Roche 2000; 
B\'ejar et al.\ 2001; Mart\'in et al.\ 2001b; Lucas et al.\ 2001), 
with the masses of some objects estimated 
to be as low as 3 Jupiter-masses (Zapatero Osorio et al.\ 2002a;
McCaughrean 2003).  However, the observational uncertainties are
still large and a cutoff in the mass function has yet to be detected.

\section{Computational method}

The calculation presented here was performed using a three-dimensional, 
smoothed particle hydrodynamics (SPH) code.  The SPH code is 
based on a version originally developed by Benz 
(Benz 1990; Benz et al.\ 1990).
The smoothing lengths of particles are variable in 
time and space, subject to the constraint that the number 
of neighbours for each particle must remain approximately 
constant at $N_{\rm neigh}=50$.  The SPH equations are 
integrated using a second-order Runge-Kutta-Fehlberg 
integrator with individual time steps for each particle 
\cite{BatBonPri1995}.
Gravitational forces between particles and a particle's 
nearest neighbours are calculated using a binary tree.  
We use the standard form of artificial viscosity 
(Monaghan \& Gingold 1983; Monaghan 1992) with strength 
parameters $\alpha_{\rm_v}=1$ and $\beta_{\rm v}=2$.
Further details can be found in Bate et al.\ \shortcite{BatBonPri1995}.
The code has been parallelised by M.\ Bate using OpenMP.  

\begin{figure}
\centerline{\psfig{figure=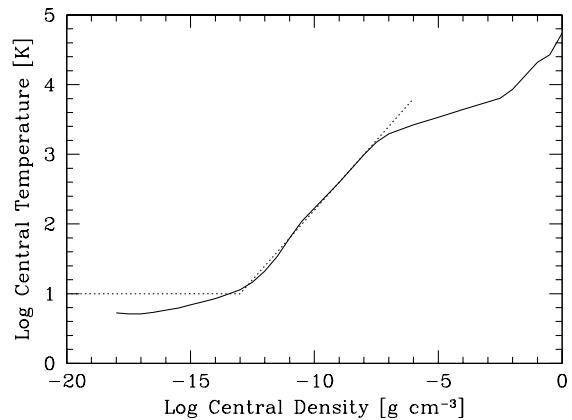,width=8.0truecm,height=5.77truecm,rwidth=8.0truecm,rheight=5.77truecm}}
\caption{\label{eos} Comparison of our barotropic equation of state (dotted line) with the temperature-density relation during the spherically-symmetric collapse of a molecular cloud core as calculated with frequency-dependent radiative transfer (solid line; Masunaga \& Inutsuka 2000).  The curves differ for densities less than $10^{-14}$ g cm$^{-3}$ simply because Masunaga \& Inutsuka chose parameters such that their initial core had a temperature of 5 K rather than our assumption of 10 K.  However, in the non-isothermal regime, from $10^{-13}$ to $10^{-8}$, our parameterisation matches the radiative transfer result to an accuracy of better than 20 percent.  The second collapse (discussed in Section 2) occurs from densities of $\approx 5\times 10^{-8}$ to $\approx 3\times 10^{-3}$ and is not modelled.  We insert a sink particle when the gas density exceeds $10^{-11}$ g cm$^{-3}$ (i.e.\ temperature $\approx 60$ K).}
\end{figure}

\subsection{Equation of state}
\label{eossec}

To model the opacity limit for fragmentation, discussed in Section 2,
without performing radiative transfer,
we use a barotropic equation of state for the thermal pressure of the
gas $p = K \rho^{\eta}$, where $K$ is a measure of the entropy
of the gas.  The value of the effective polytropic exponent $\eta$, 
varies with density as
\begin{equation}\label{eta}
\eta = \cases{\begin{array}{rl}
1, & \rho \leq 10^{-13}~ {\rm g~cm}^{-3}, \cr
7/5, & \rho > 10^{-13}~ {\rm g~cm}^{-3}. \cr
\end{array}}
\end{equation}
We take the mean molecular weight of the gas to be $\mu = 2.46$.
The value of $K$ is defined such that when the gas is 
isothermal $K=c_{\rm s}^2$, with the sound speed
$c_{\rm s} = 1.84 \times 10^4$ cm s$^{-1}$ at 10 K,
and the pressure is continuous when the value of $\eta$ changes.

This equation of state has been chosen to match closely the 
relationship between temperature and density during the 
spherically-symmetric collapse of molecular 
cloud cores as calculated with frequency-dependent radiative 
transfer (Masunaga, Miyama, \& Inutsuka 1998; 
Masunaga \& Inutsuka 2000).  A comparison of our simple 
parameterisation with Masunaga and Inutsuka's temperature-density 
relation is given in Figure \ref{eos}.  Our parameterisation
reproduces the temperature-density relation to an accuracy
of better than 20\% in the non-isothermal regime up to densities
of $10^{-8}$ g~cm$^{-3}$.
Test calculations of the spherically-symmetric collapse of a 
molecular cloud core using this equation of state have been performed 
(Bate 1998, 2002) and give excellent agreement with the 
results of Larson \shortcite{Larson1969} and 
Winkler \& Newman (1980a,b) for the mass and size of the 
pressure-supported fragment that forms.
Thus, our equation of state should model collapsing 
regions well, but may not model the equation of 
state in protostellar discs particularly accurately due to
the departure from spherical symmetry.

\subsection{Sink particles}
\label{sinkparticles}

The opacity limit for fragmentation results in the formation 
of distinct pressure-supported
fragments in the calculation.  As these fragments accrete, their
central density increases, and it becomes computationally impractical
to follow their internal evolution until they undergo the second
collapse to form stellar cores because of the short dynamical
time-scales involved.  Therefore, when the central density of 
a pressure-supported fragment exceeds 
$\rho_{\rm s} = 10^{-11}~{\rm g~cm}^{-3}$, 
we insert a sink particle into the calculation 
Bate et al.\ \shortcite{BatBonPri1995}.

In the calculation presented here, a sink particle is formed by 
replacing the SPH gas particles contained within $r_{\rm acc}=5$ AU 
of the densest gas particle in a pressure-supported fragment 
by a point mass with the same mass and momentum.  Any gas that 
later falls within this radius is accreted by the point mass 
if it is bound and its specific angular momentum is less than 
that required to form a circular orbit at radius $r_{\rm acc}$ 
from the sink particle.  Thus, gaseous discs around sink 
particles can only be resolved if they have radii $\gsim 10$ AU.
Sink particles interact with the gas only via gravity and accretion.

Since all sink particles are created from pressure-supported 
fragments, their initial masses are $\approx 10$ M$_{\rm J}$, 
as given by the opacity limit for fragmentation (Section \ref{opacitysec}).  
Subsequently, they may accrete large amounts of material 
to become higher-mass brown dwarfs ($\lsim 75$ M$_{\rm J}$) or 
stars ($\gsim 75$ M$_{\rm J}$), but {\it all} the stars and brown
dwarfs begin as these low-mass pressure-supported fragments.

The gravitational acceleration between two sink particles is
Newtonian for $r\geq 4$ AU, but is softened within this radius
using spline softening \cite{Benz1990}.  The maximum acceleration 
occurs at a distance of $\approx 1$ AU; therefore, this is the
minimum separation that a binary can have even if, in reality,
the binary's orbit would have been hardened.  Sink particles are
not permitted to merge in this calculation.

Replacing the pressure-supported fragments with sink particles 
is necessary in order to perform the calculation.  However, it is 
not without an element of risk.  If it were possible to follow the 
fragments all the way to stellar densities (as done by Bate 1998)
while continuing to follow the evolution of the large-scale cloud over its 
dynamical time-scale, we might find that a few of the objects that we 
replace with sink particles merge together or are
disrupted by dynamical interactions.  We have tried to minimise the 
degree to which this might occur by insisting that the central
density of the pressure-supported fragments exceeds $\rho_{\rm s}$
before a sink particle is created.  This is two orders of magnitude 
higher than the density at which the gas is heated and ensures that
the fragment is self-gravitating, centrally-condensed and, in practice, 
roughly spherical before it 
is replaced by a sink particle (see, for example, the fragments
that have not yet been replaced by sink particles in Figure \ref{core2} 
at $t=1.337$).  In theory, it would be possible 
for a long collapsing filament to exceed this density over a large
distance, thus making the creation of one or more sink 
particles ambiguous.  However, the structure of the collapsing 
gas that results 
from the turbulence prohibits this from occurring; no long, roughly 
uniform-density filaments with densities $\approx \rho_{\rm s}$
form during the calculation.  Furthermore, each pressure-supported 
fragment must undergo a period of accretion before its central density
exceeds $\rho_{\rm s}$ and it is replaced by a sink particle.  
For example, it is common in the calculation
to be able to follow a pressure-supported fragment that forms via 
gravitational instability in a disc for roughly half an orbital 
period before it is replaced.  Thus, the fragments do have some time 
in which they may merge or be disrupted.  
Only occasionally are low-mass pressure supported fragments 
disrupted during the calculation; most are eventually 
replaced by sink particles.

\begin{table*}
\begin{tabular}{lccccccc}\hline
Core & Initial Gas  & Initial  & Final  & No. Stars & No. Brown  & Mass of Stars and  & Star Formation \\
 & Mass  & Size & Gas Mass & Formed & Dwarfs Formed & Brown Dwarfs & Efficiency  \\
 & M$_\odot$ & pc & M$_\odot$ &  & & M$_\odot$ & \% \\ \hline
1 & 3.00 (0.76) & $0.06\times 0.04\times 0.03$ & 3.66 (1.59) & $\geq$17 & $\leq$21 & 4.96 & 58 (76) \\
2\&3 & 1.80 (0.42) & $0.08\times 0.02\times 0.02$ & 2.09 (0.55) & $\geq$6 & $\leq$6 & 0.93 & 31 (63)\\
2 & 0.88 (0.21) & ($0.02\times 0.01\times 0.01$) & 1.00 (0.24) & $\geq$3 & $\leq$4 & 0.47 & 32 (66)\\
3 & 1.10 (0.32) & ($0.02\times 0.01\times 0.01$) & 1.09 (0.32) & $\geq$3 & $\leq$2 & 0.46 & 30 (59)\\ \hline
Cloud & 50.0 & $0.38\times 0.38\times 0.38$ & 44.1 & $\geq$23 & $\leq$27 & 5.89 & 12 \\ \hline
\end{tabular}
\caption{\label{table1} The properties of the three dense cores that form during the calculation and those of the cloud as a whole.  The gas masses and sizes of the cores are calculated from gas with $n({\rm H}_2)>1\times 10^6$~cm$^{-3}$ and $n({\rm H}_2)>1\times 10^7$~cm$^{-3}$ (the latter values are given in parentheses).  We note that the gas mass of a core depends on the density above which gas is included in the calculation as $M\propto \rho^{-\gamma}$ with $\gamma\approx 0.5$.  Cores 2 \& 3 are joined when the lower density criterion is used.  The initial gas mass is calculated just before star formation begins in that core (i.e.\ different times for each core).  Brown dwarfs have final masses less than 0.075 M$_\odot$.  The star formation efficiency is taken to be the total mass of the stars and brown dwarfs that formed in the core divided by the sum of this mass and the mass in gas in the core at the end of the calculation.  Note that the star formation efficiency is high locally, but low globally.  The numbers of stars (brown dwarfs) are lower (upper) limits because nine of the brown dwarfs are still accreting when the calculation is stopped (Section 4). }
\end{table*}

\subsection{Initial conditions}

The initial conditions consist of a large-scale, turbulent 
molecular cloud.  The cloud is spherical and
uniform in density with mass of 50 M$_\odot$ and
a diameter of $0.375$ pc (77400 AU).  At the temperature of 10 K,
the mean thermal Jeans mass is 1 M$_\odot$ 
(i.e.\ the cloud contains 50 thermal Jeans masses).
The free-fall time of the cloud is $t_{\rm ff}=6.0\times 10^{12}$~s
or $1.90\times 10^5$ years.

Although the cloud is uniform in density, we impose an initial 
supersonic turbulent velocity field on it in the same manner
as Ostriker, Stone \& Gammie \shortcite{OstStoGam2001}.  We generate a
divergence-free random Gaussian velocity field with a power spectrum 
$P(k) \propto k^{-4}$, where $k$ is the wavenumber.  
In three dimensions, this results in a
velocity dispersion that varies with distance, $\lambda$, 
as $\sigma(\lambda) \propto \lambda^{1/2}$ in agreement with the 
observed Larson scaling relations for molecular clouds 
\cite{Larson1981}.
This power spectrum is slightly steeper than the Kolmogorov
spectrum, $P(k)\propto k^{-11/3}$.  Rather, it matches the 
amplitude scaling of Burgers supersonic turbulence associated
with an ensemble of shocks (but differs from Burgers turbulence
in that the initial phases are uncorrelated).
The velocity field is generated on a $64^3$ uniform grid and the
velocities of the particles are interpolated from the grid.  The
velocity field is normalised so that the kinetic energy of the 
turbulence equals the magnitude of the gravitational potential 
energy of the cloud (i.e.\ initially, the cloud has more than enough
turbulent energy to support itself against gravity).
The initial root-mean-square Mach number of the turbulence 
is ${\cal M}=6.4$.

\subsection{Resolution}

The local Jeans mass must be resolved throughout 
the calculation to model fragmentation correctly 
(Bate \& Burkert 1997; Truelove et al.\ 1997; Whitworth 1998; 
Boss et al.\ 2000).  Bate \& Burkert \shortcite{BatBur1997} 
found that this requires 
$\gsim 2 N_{\rm neigh}$ SPH particles per Jeans mass; 
$N_{\rm neigh}$ is insufficient.
We have repeated their calculation using different numbers of
particles and find that $1.5 N_{\rm neigh}=75$
particles is also sufficient
to resolve fragmentation (see Appendix A).
The minimum Jeans mass in the calculation presented here occurs 
at the maximum density during the isothermal phase of the 
collapse, $\rho = 10^{-13}$ g~cm$^{-3}$, 
and is $\approx 0.0011$ M$_\odot$ (1.1 M$_{\rm J}$).  Thus, we
use $3.5 \times 10^6$ particles to model the 50-M$_\odot$ cloud.
This SPH calculation is one of the largest ever performed.
It required approximately 95000 CPU hours on the SGI Origin 3800 
of the United Kingdom Astrophysical Fluids Facility (UKAFF).

\section{The evolution of the cloud}
\label{evolution}

\begin{figure*}
\centerline{\psfig{figure=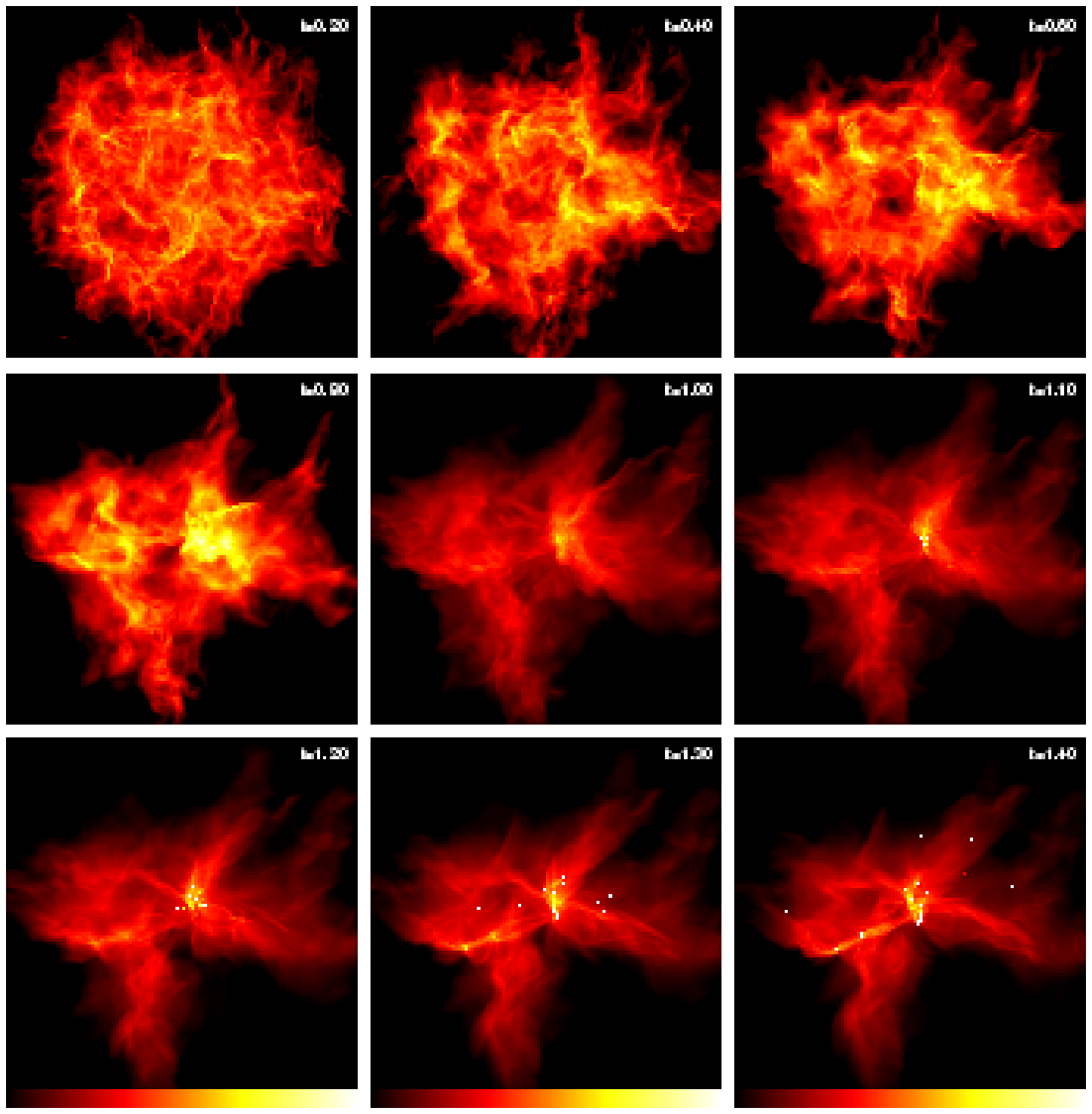,width=17.5truecm}}
\caption{\label{cloud} The global evolution of the cloud during the calculation. Shocks lead to dissipation of the turbulent energy that initially supports the cloud, allowing parts of the cloud to collapse.  Star formation begins at $t=1.04 t_{\rm ff}$ in a collapsing dense core.  By the end of the calculation, two more dense cores have begun forming stars (lower left of the last panel) and many of the stars and brown dwarfs have been ejected from the cloud through dynamical interactions.  Each panel is 0.4 pc (82400 AU) across.  Time is given in units of the initial free-fall time of $1.90\times 10^5$ years.  The panels show the logarithm of column density, $N$, through the cloud, with the scale covering $-1.5 < \log N < 0$ for $t<1.0$ and $-1.7 < \log N < 1.5$ for $t\geq 1.0$ with $N$ measured in g cm$^{-2}$.}
\end{figure*}

\begin{figure*}
\centerline{\psfig{figure=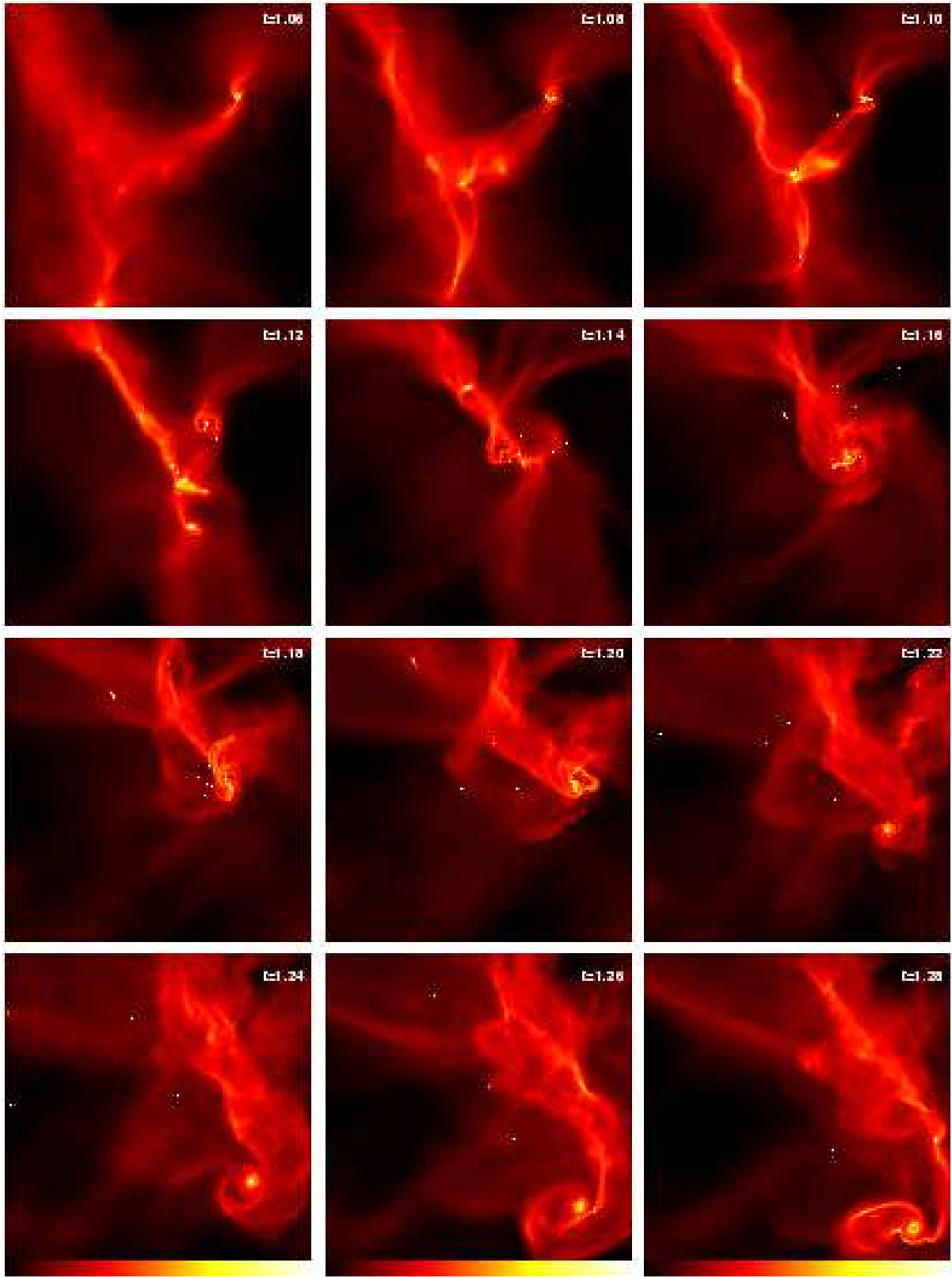,width=16.0truecm}}
\caption{\label{core1a} The star formation in the first (main) dense core.  The first objects form a binary at $t=1.037 t_{\rm ff}$.  Large gaseous filaments collapse to form single objects and multiple systems.  These objects fall together to form a small group.  The group quickly dissolves due to dynamical interactions and, simultaneously, there is a quiet period ($t=1.16-1.24 t_{\rm ff}$) in the star formation while more gas falls into the core.  At $t\approx 1.26$, a new burst of star formation begins in the filamentary gas and in a large disc around a close binary.  The sequence is continued in Figure 4.  Each panel is 0.025 pc (5150 AU) across.  Time is given in units of the initial free-fall time of $1.90\times 10^5$ years.  The panels show the logarithm of column density, $N$, through the cloud, with the scale covering $-0.5 < \log N < 2.5$ with $N$ measured in g cm$^{-2}$.}
\end{figure*}

\begin{figure*}
\centerline{\psfig{figure=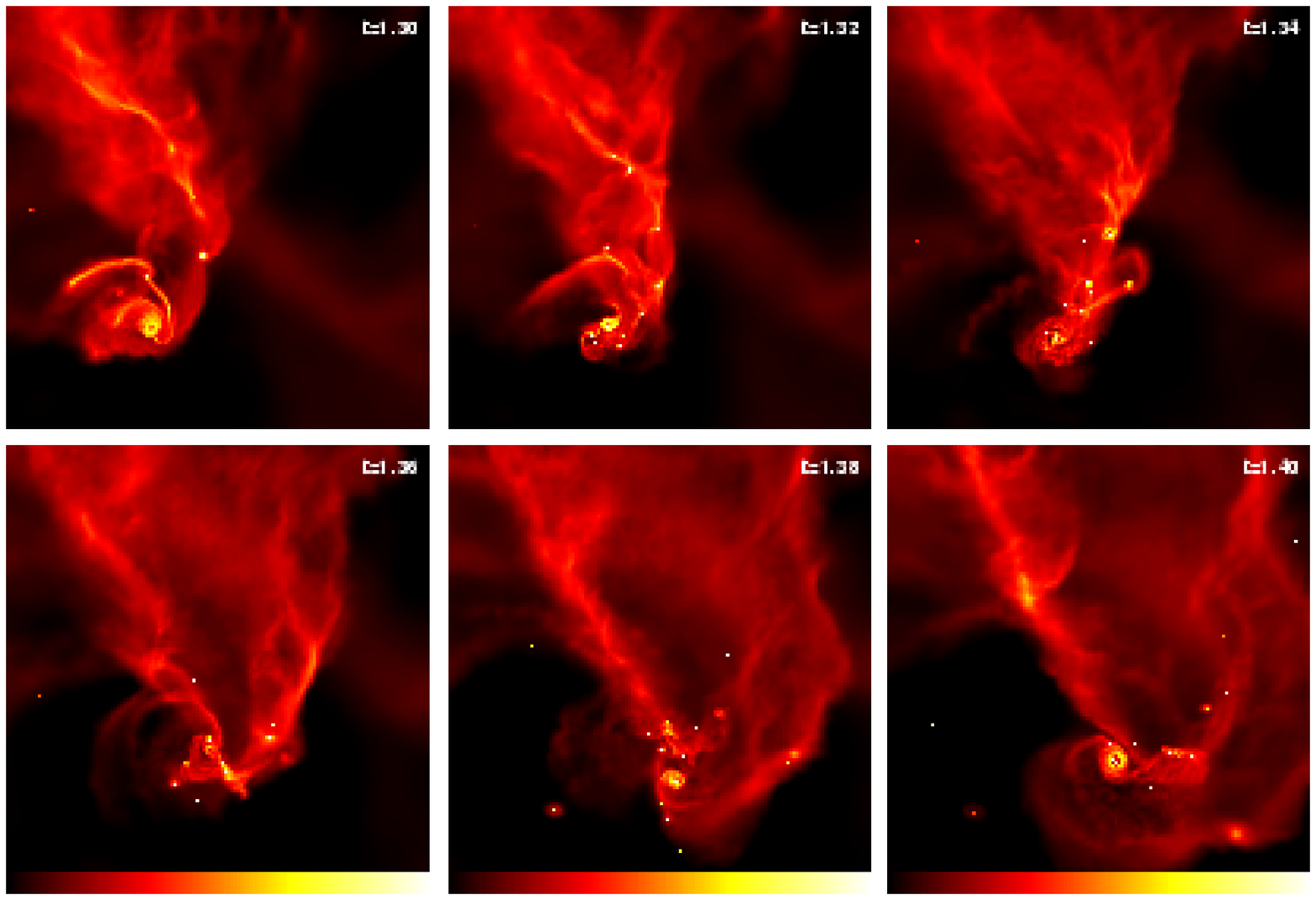,width=16.0truecm}}
\caption{\label{core1b} The star formation in the first (main) dense core, continued from Figure 3.  The second burst of star formation again produces a small group of objects.  This group has almost dissolved by the time the calculation is stopped.  Notable events include the ejection of a brown dwarf with a resolved disc ($t=1.38 t_{\rm ff}$, lower left).  Each panel is 0.025 pc (5150 AU) across.  Time is given in units of the initial free-fall time of $1.90\times 10^5$ years.  The panels show the logarithm of column density, $N$, through the cloud, with the scale covering $-0.5 < \log N < 2.5$ with $N$ measured in g cm$^{-2}$. }
\end{figure*}

Although the initial turbulent velocity field is divergence-free,
hydrodynamic evolution of the cloud soon results in the formation
of shocks.  Small-scale shocks form first, followed by larger 
structures as the calculation proceeds 
(Figure \ref{cloud}, $t=0-0.8$ t$_{\rm ff}$).  
Kinetic energy is lost from the cloud in these shocks.  
Therefore, although the cloud initially has more than enough turbulent
energy to support itself, gravity soon begins to dominate and collapse
occurs in parts of the cloud.  The turbulence decays on the dynamical
time-scale of the cloud, and star formation begins after just 
one global free-fall time at $t=1.037~t_{\rm ff}$ 
(i.e.\ $t=1.97\times 10^5$ yrs).  By this time, the root-mean-square 
Mach number of the turbulence has fallen from its initial value
of ${\cal M}=6.4$ to only ${\cal M}=3.8$.  
This rapid decay of the turbulence
is consistent with other numerical studies of turbulence in 
molecular clouds, both with and without magnetic fields and 
self-gravity (e.g.\ MacLow et al.\ 1998; Stone et al.\ 1998; 
Ostriker et al.\ 2001).  

Regions of overdensity form from converging gas flows during the
decay of the turbulence collapse  
(Figure \ref{cloud}, $t\geq 0.8$ t$_{\rm ff}$), producing three 
dense star-forming cores within the cloud 
(Figure \ref{cloud}, $t=1.0$ and $t=1.2$ t$_{\rm ff}$; Figures 
\ref{core1a}, \ref{core1b}, \ref{core2}, \ref{core3}).  
The properties of these cores and the numbers of 
stars and brown dwarfs they produce during the calculation 
are summarised in Table 1.  Figure \ref{sfrate} gives the 
times at which each 
star or brown dwarf is formed (i.e.\ the time at which the 
pressure-supported fragments are replaced by sink particles) 
in terms of the number of free-fall times from 
the start of the calculation (and also in years).

The most massive core begins 
forming stars first (Figures \ref{core1a}, \ref{core1b}).  
It contains $\approx 3.0$ M$_\odot$ 
when star formation begins ($\approx 6$\% of the mass 
of the cloud), although this figure depends on the density threshold 
that is used (Table 1).  
The core undergoes an initial burst of star 
formation lasting $\approx 18000$ years followed 
by a quiet period of $\approx 24000$ years during which only 
three brown dwarfs form (see Figure \ref{sfrate}).  
This pattern then repeats with another
burst lasting $\approx 18000$ years and a quiet period that lasts
the remaining $\approx 9000$ years until the end of the calculation,
during which time only one brown dwarf forms in the most massive core.  
At this point we had exhausted our allocation of computer time and
the calculation was stopped at $t=1.40$ t$_{\rm ff}$
($t=2.66\times 10^5$ years).  In all, 
38 stars and brown dwarfs formed in the most massive core.

The second and third dense cores begin forming stars at 
$t=1.296$ t$_{\rm ff}$ ($t=2.47\times 10^5$ years) and 
$t=1.318$ t$_{\rm ff}$ ($t=2.51\times 10^5$ years), respectively.  
They each contain around 1 M$_\odot$
of gas when the star formation begins.  They produce
7 and 5, respectively, objects during the calculation.

The evolution of the velocity field and the density structure in
the cloud will be discussed in more detail in a later paper.  
In this paper, we concentrate on the process of star formation 
and the properties of the resulting stars and brown dwarfs.  
When the calculation is stopped, the cloud has produced 23 stars
and 18 brown dwarfs.  An additional 9 objects have substellar masses
but are still accreting.  Three of these have very low masses and
accretion rates and therefore would probably end up
with substellar masses if the calculation were continued.  The
other six already have masses near the stellar/substellar boundary 
and are therefore likely to become stars.

\subsection{The star formation process in the dense cores}

\begin{figure*}
\centerline{\psfig{figure=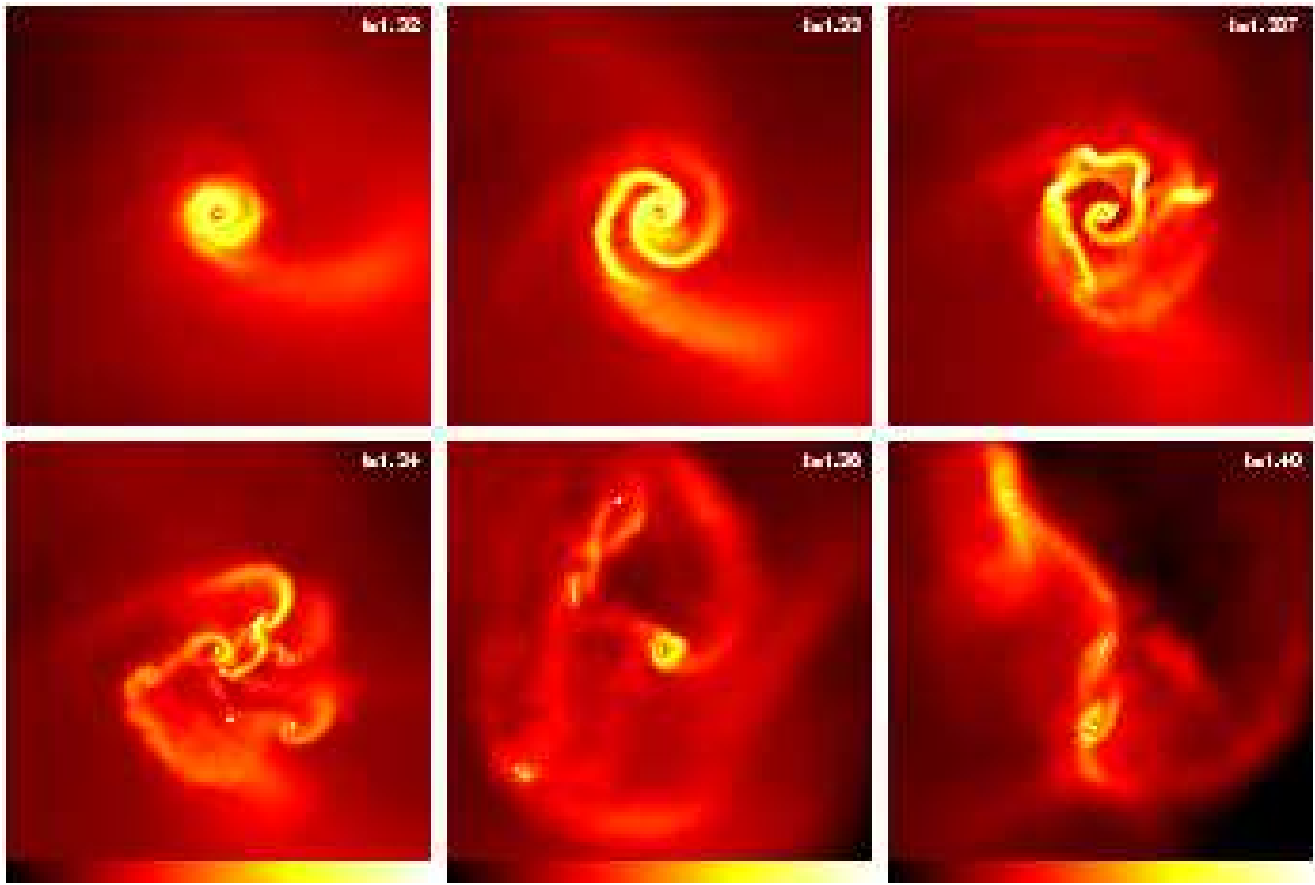,width=17.5truecm}}
\caption{\label{core2} The star formation in the second dense core.  The first object forms at $t=1.296 t_{\rm ff}$, and a circumstellar disc forms around it.  During the 800-year period between $t=1.334$ and $t=1.338 t_{\rm ff}$, the circumstellar disc (now very massive) fragments to form six more objects.  These objects undergo chaotic interactions and, at the end of the calculation, the system is composed of an unstable triple system orbiting a quadruple system which is itself composed of two close binary systems.  Each panel is 600 AU across.  Time is given in units of the initial free-fall time of $1.90\times 10^5$ years.  The panels show the logarithm of column density, $N$, through the cloud, with the scale covering $0.0 < \log N < 2.5$ with $N$ measured in g cm$^{-2}$.}
\end{figure*}

\begin{figure*}
\centerline{\psfig{figure=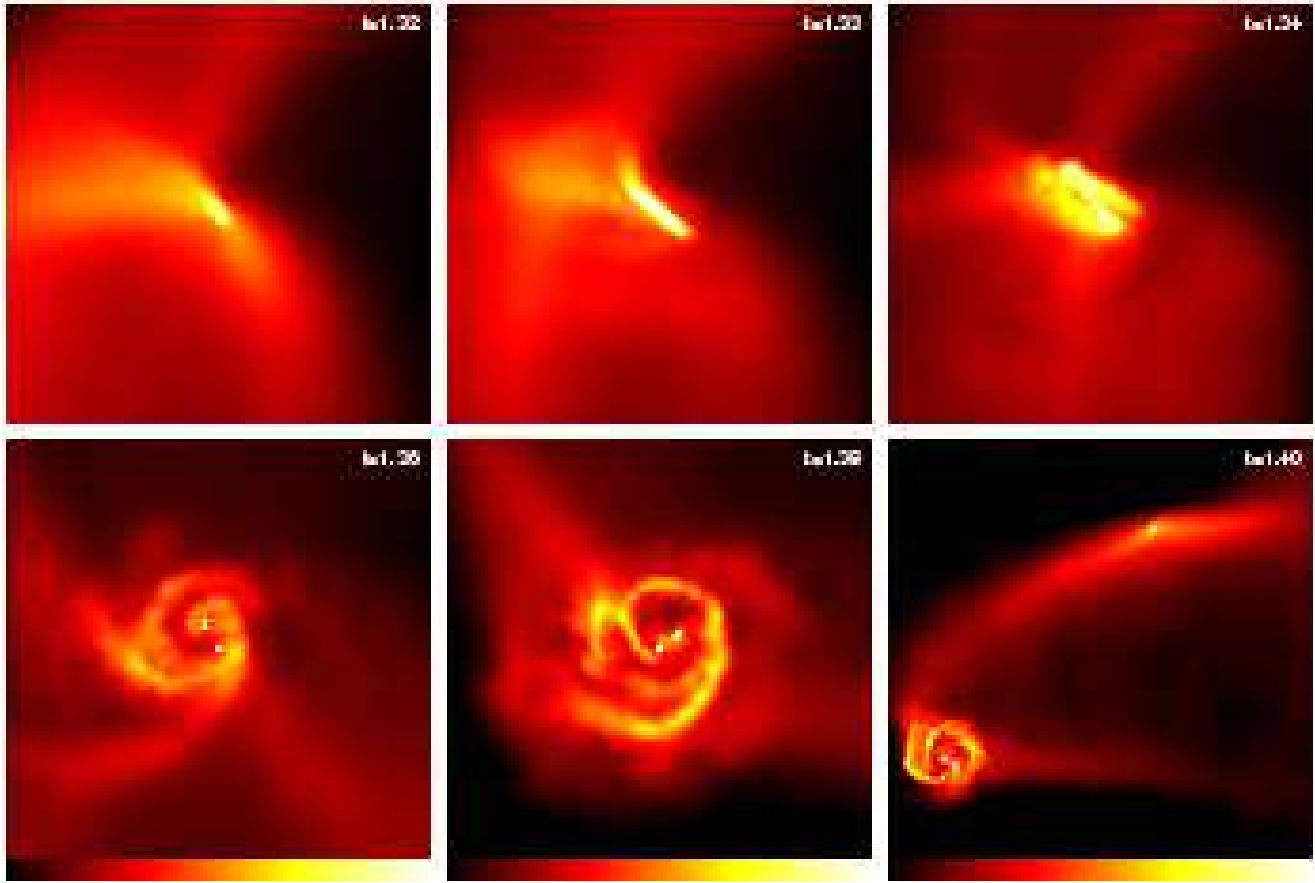,width=17.5truecm}}
\caption{\label{core3} The star formation in the third dense core.  The first object forms at $t=1.318 t_{\rm ff}$, and an edge-on circumstellar disc forms around it.  At $t=1.338 t_{\rm ff}$, the circumstellar disc (now very massive) fragments to form two more objects.  Just before the calculation is stopped, a fourth object fragments out of the circumtriple disc and a fifth object forms in a nearby filament.  Each panel is 400 AU across except the last which is 1000 AU across.  Time is given in units of the initial free-fall time of $1.90\times 10^5$ years.  The panels show the logarithm of column density, $N$, through the cloud, with the scale covering $0.0 < \log N < 2.5$ with $N$ measured in g cm$^{-2}$.}
\end{figure*}

Snapshots of the process of star formation in the most massive core
are shown in Figures \ref{core1a} and \ref{core1b}.  The star 
formation in the two low-mass cores is depicted in Figures 
\ref{core2} and \ref{core3}.  A true appreciation of how dynamic 
and chaotic the star formation process is can only be obtained by 
studying an animation of the simulation.  The reader is
encouraged to download an animation of the simulation from one of the
two internet sites 
http://www.astro.ex.ac.uk/people/mbate/Research/Cluster/
or http://www.ukaff.ac.uk/starcluster .

The gravitational collapse of the most massive dense core produces
filamentary structures which fragment (e.g.\ Bastein 1983; 
Bastien et al.\ 1991; Inutsuka \& Miyama 1992) to form a combination of 
single objects and multiple systems (Figure \ref{core1a}, 
$t=1.06-1.10$ t$_{\rm ff}$).  Many of the multiple systems result
from the fragmentation of massive circumstellar discs that form 
around single objects that fragment out of the filaments
(e.g.\ Bonnell 1994; Bate \& Bonnell 1994; Whitworth et al.\ 1995;
Burkert, Bate \& Bodenheimer 1997).  Subsequently, most of these 
objects fall together into the gravitational potential well to form
a small stellar cluster 
(Figure \ref{core1a}, $t=1.12-1.14$ t$_{\rm ff}$).
The cluster only contains $\approx 13$ objects and, thus, 
dynamical interactions quickly result in its dissolution 
(Figure \ref{core1a}, $t=1.16-1.22$ t$_{\rm ff}$).
The formation of stars is essentially halted during this dissolution
phase because a significant fraction of the gas was used up in 
producing the stars and brown dwarfs.  Later, when more gas has 
fallen into the potential well of the main dense core, a new 
burst of star formation occurs
(Figures \ref{core1a} and \ref{core1b}, $t=1.26-1.34$ t$_{\rm ff}$) 
producing a second cluster
that contains $\approx 16$ objects at any one time.  Again, this
small cluster quickly disperses due to dynamical interactions 
(Figure \ref{core1b}, $t=1.34-1.40$ t$_{\rm ff}$), although at 
the end of the calculation it has not yet fully dissolved
and still contains $\approx 11$ objects.

The second and third dense cores produce only a small 
number of objects compared to the most massive core (Table 1).  
The dynamical interactions in these two stellar groups work 
to arrange dynamically unstable systems into more stable configurations
(Figures \ref{core2} and \ref{core3}).  
During the calculation, neither of these low-mass cores ejects
any objects, although ejections are likely in the long-term.  
The formation of the multiple systems in these low-mass cores 
results almost exclusively from the fragmentation of massive 
circumstellar discs (Figures \ref{core2} and \ref{core3}).  
Only in the third core, near the end of the calculation, is an 
object formed in a separate filament (Figure \ref{core3}, 
$t=1.40$ t$_{\rm ff}$).

\begin{figure*}
\centerline{\psfig{figure=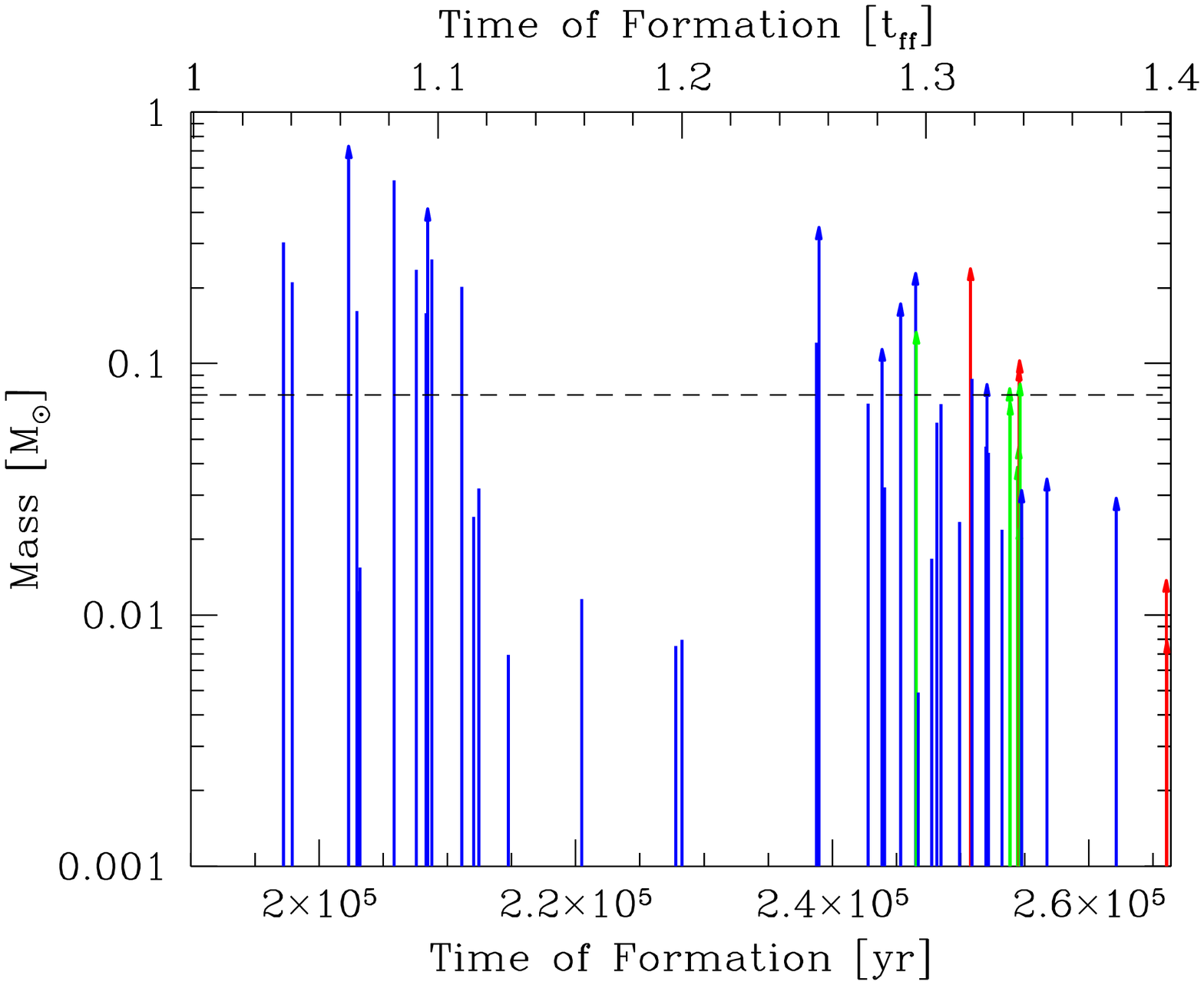,width=15.0truecm,height=12.125truecm,rwidth=15.0truecm,rheight=12.0truecm}}
\caption{\label{sfrate} Time of formation and mass of each star and brown dwarf at the end of the calculation.  The colour of each line identifies the dense core in which the object formed: first (blue), second (green), or third (red) core.  Objects that are still accreting significantly at the end of the calculation are represented with arrows.  The horizontal dashed line marks the star/brown dwarf boundary.  Time is measured from the beginning of the calculation in terms of the free-fall time of the initial cloud (top) or years (bottom).  The star formation occurs in two bursts that each last $\approx 2\times 10^4$ years.}
\end{figure*}

\begin{figure}
\centerline{\psfig{figure=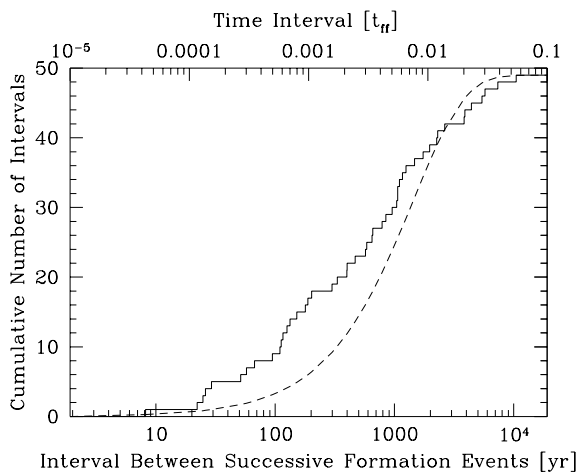,width=8.0truecm,height=8.0truecm,rwidth=8.0truecm,rheight=6.5truecm}}
\caption{\label{poisson} The cumulative distribution of the time intervals between successive star formation events in Figure 7 (solid line) compared with a Poisson distribution (dashed line).  There is an excess of short time intervals between star formation events.  The time intervals are given in units of the free-fall time of the initial cloud (top) or years (bottom).}
\end{figure}

The formation of the stars and brown dwarfs is not a Poisson process.
In Figure \ref{sfrate}, along with the two overall bursts of star
formation, it appears to be common for pairs of objects or several 
objects to be formed 
within a short time of each other.  This can be investigated by 
considering the distribution of time intervals between successive 
star formation events.  In Figure \ref{poisson}, we plot the 
cumulative distribution of these time intervals (solid line) 
and compare it with the distribution that would be expected 
if the events were Poissonian (dashed line).  A Kolmogorov-Smirnov
test of the two distributions gives only a 0.7\% probability of 
the time intervals being distributed in a Poisson manner.  Instead,
there is an excess of short time intervals.
This is due to multiple fragmentation events in 
gravitationally-unstable discs 
(e.g.\ Figure \ref{core2}, $t=1.33-1.34$ t$_{\rm ff}$), along with the 
fragmentation of collapsing gas filaments to form binaries 
(e.g.\ the first two stars to form, caption of Figure \ref{core1a}).
Thus, in addition to the bursts of star formation on the scales of
molecular cloud cores, fragmentation events on small-scales 
are correlated.

\subsection{Comparison with observed star-forming regions}
\label{compare}

Comparing our calculation to real star-forming regions that have been 
well studied, this calculation could be regarded as modelling
part of the $\rho$ Ophiuchus dark cloud.  The main cloud of
Ophiuchus contains around 550 M$_\odot$ of gas within an 
area on the sky of around $2\times 1$ pc (Wilking \& Lada 1983).  
It is centrally-condensed and contains 6 main dense cores.
The gas masses of the cores range from $\approx 8-62$ M$_\odot$
with mean densities of $n({\rm H_2})\sim 10^5-10^6$ cm$^{-3}$ 
(Motte, Andre \& Neri 1998).  The calculation
presented here contains a total of 50 M$_\odot$ and 
forms one large and two smaller dense cores
in a region $\approx 0.4$ pc across.  Just before the star formation 
begins, the large core contains $\approx 5.3$ M$_\odot$ and measures 
$\approx 0.1$ pc in diameter (counting gas
with $n({\rm H_2})>3\times 10^5$~cm$^{-3}$).  Its mass, mean density, 
and size are similar
to those of the Ophiuchus-F core which contains $8$ M$_\odot$
(Motte et al.\ 1998).
Thus, the calculation presented here is 
comparable to modelling a region containing one of the 
main dense cores within the $\rho$ Ophiuchus dark cloud.

Alternatively, the calculation can be viewed as modelling part
of the cloud that formed the Orion Trapezium Cluster.  Although
the total mass of our cloud is far less than the progenitor
of the Trapezium Cluster and our cloud produces only low-mass stars,
it is similar in terms of the resulting stellar density.  The
stellar densities in the centre of the Trapezium Cluster
are $2 - 4\times 10^4$ pc$^{-3}$ (McCaughrean \& Stauffer 1994; Hillenbrand
\& Hartmann 1998; Bate, Clarke \& McCaughrean 1998) and the
density falls off with radius as 
$n\approx 2\times 10^4 (r/0.07~{\rm pc})^{-2}$ pc$^{-3}$
(Bate et al.\ 1998).  Our calculation produces $\approx 30$ {\it stars}
from a cloud with
an initial diameter of 0.375 pc giving a stellar density of 
$\approx 1\times 10^3$ pc$^{-3}$.  Thus, the overall stellar densities in 
our calculation are similar to those in the Trapezium Cluster
0.3 pc from the centre (a calculation of the radius at which the
stellar densities fall to $10^3$ pc$^{-3}$ using Hillenbrand \&
Hartmann's King model fit with $r_0=0.16$ pc gives a radius of
$\approx 0.38$ pc).  This is within the Trapezium Cluster's
half-mass radius of 0.8 pc (Hillenbrand \& Hartmann 1998).  
Thus, for processes that depend on stellar densities, such 
as the dynamical truncation of circumstellar discs, our 
calculation can also be compared with the Orion Trapezium Cluster.

\section{Properties of the stars and brown dwarfs}
\label{properties}

For the remainder of the paper, we examine the properties of the
stars and brown dwarfs that form in the calculation and compare them
to the observed properties of stars and brown dwarfs.  In this way,
we aim to test whether our current understanding of the 
fragmentation of a turbulent molecular cloud is realistic and to
predict some properties of stars and brown dwarfs that have
not yet been determined observationally.

\subsection{Star formation efficiency and timescale}
\label{efficiencytimescale}

As described above, the star formation occurs in one large and 
two small dense cores that form within the initial cloud due to the
decay of the initial supersonic turbulence.  The properties 
of the cores and the stars and brown dwarfs they produce are 
given in Table 1, along with the totals for the cloud as a whole.
For each of the cores, the local star formation efficiency is high.
In fact, it is this high efficiency that is responsible for the bursts of 
star formation in Figure \ref{sfrate}.  The most massive core undergoes
a rapid burst of star formation 
(Figure \ref{core1a}, $t=1.06-1.12$ t$_{\rm ff}$)
lasting $\approx 18000$ years that severely depletes its 
reservoir of gas, temporarily halting the star formation.
However, the core (including those stars and
brown dwarfs that are not rapidly ejected) still 
dominates the local gravitational potential and, therefore, attracts
more gas (Figure \ref{core1a}, $t=1.14-1.24$ t$_{\rm ff}$).  
When the gas becomes sufficiently dense, a new
burst of star formation occurs
(Figure \ref{core1a}, $t=1.26-1.34$ t$_{\rm ff}$) 
and the process is repeated.

Although the local star formation efficiency is high in the dense
cores, most of the gas in the cloud is in low-density regions where 
no star formation occurs.  Thus, the overall star formation efficiency
of the cloud is low, $\approx 10$\%.  Due to computational 
limitations, we have not been able to follow the cloud until
star formation ceases entirely.  However, by the end of the
calculation a large fraction of the initial cloud has drifted
off to large distances due to a combination of the initial
velocity dispersion and pressure gradients and is not gravitationally
unstable.  Thus, the global star formation efficiency is unlikely to rise
above a few tens of percent.  Furthermore, the
calculation neglects all feedback processes.  The most massive
star to form is only 0.73 M$_\odot$ so feedback may not
be very important.  Even so, jets, outflows and heating of the 
gas may be expected to reduce the efficiency of star formation still 
further.  One way in which the star formation efficiency might be
increased is if the initial turbulence in the cloud was reduced
since then less of the gas would be able to drift away from the cloud.

Observations show that star formation efficiencies vary widely
across star-forming regions.  Efficiencies may be high locally,
but are generally low globally.  For example, in the 
$\rho$ Ophiuchus cloud, Wilking and Lada \shortcite{WilLad1983} 
find the overall star formation efficiency to be 20-30\%
but may be as high as 47\% locally.  On a grander
scale, star formation in the Orion B molecular cloud does not occur
uniformly but is concentrated in massive ($M>200$ M$_\odot$) dense 
cores that have high star formation efficiencies \cite{Lada1992}.  
Similarly, 
the Orion Trapezium Cluster has a high star formation efficiency 
\cite{HilHar1998}, but it is just one small part of the Orion A
molecular cloud.  We note that the star 
formation efficiency is much easier to measure from simulations
than from observations because stars can quickly disperse from the
dense cores in which they formed (see Section \ref{stellarveldisp}).

\begin{figure*}
\centerline{\psfig{figure=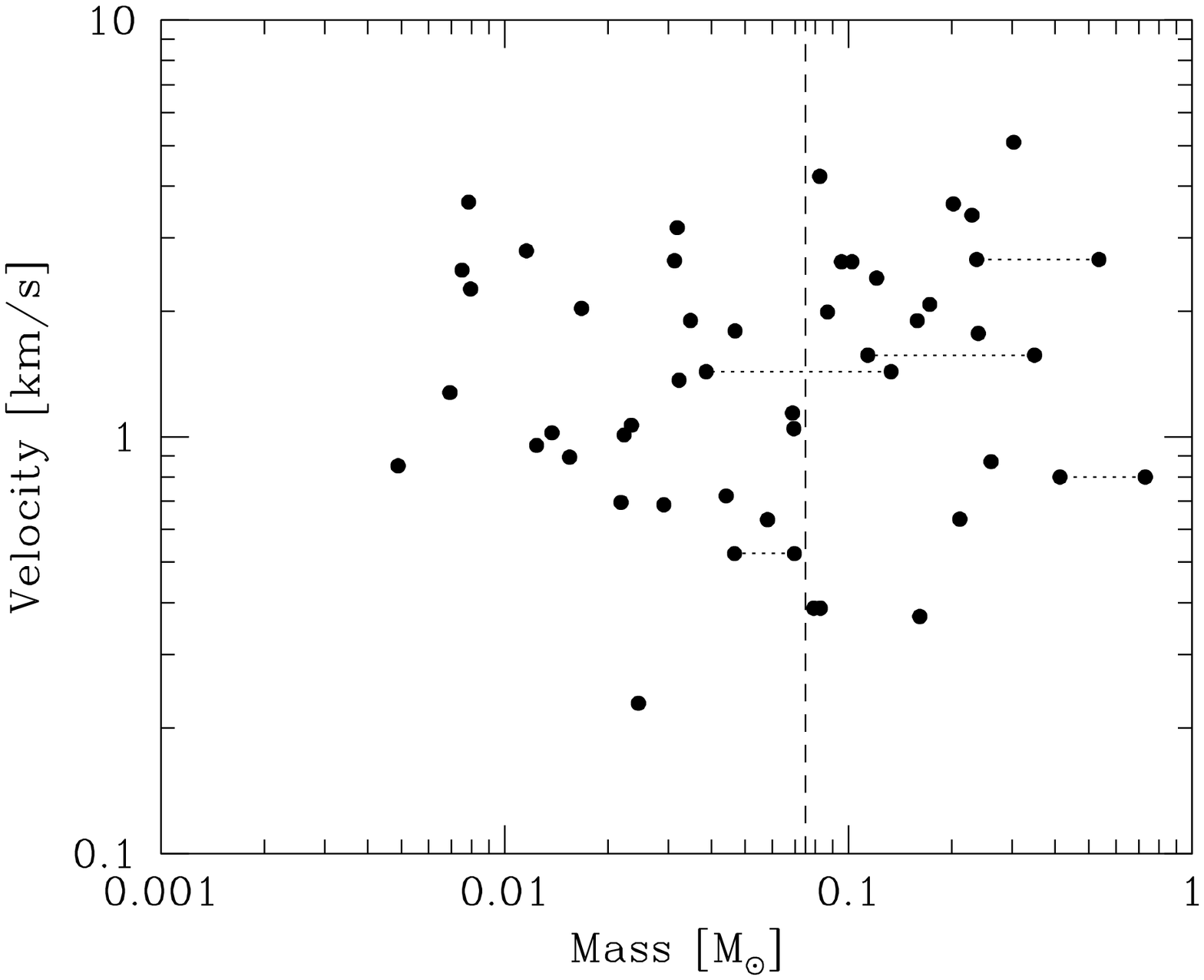,width=15.0truecm,height=12.91truecm,rwidth=15.0truecm,rheight=12.0truecm}}
\caption{\label{veldisp} The velocities of each star and brown dwarf relative to the centre-of-mass velocity of the stellar system.  For close binaries (semi-major axes $< 10$ AU), the centre-of-mass velocity of the binary is given, and the two stars are connected by dotted lines.  The root mean square velocity dispersion for the association (counting each binary once) is 2.1 km/s (3-D) or 1.2 km/s (1-D).  This is in good agreement with low-mass star-forming regions.  There is no significant dependence of the velocity dispersion on either mass or binarity.  The vertical dashed line marks the star/brown dwarf boundary.  If the centre-of-mass velocity is used for multiple systems with semi-major axes $<100$ AU and each system is counted only once, the 3-D velocity dispersion is reduced slightly to 1.8 km/s.}
\end{figure*}

The timescale on which star formation occurs is the dynamical
one.  Star formation begins
after one global free-fall time of the cloud.  Similarly, the two 
bursts of star formation in the most massive core, each lasting for 
$\approx 18000$ years, and the time between the bursts (i.e.\ the 
time-scale for gas to fall into the core and replenish it) of
$\approx 24000$ years are all roughly equal to the dynamical timescale 
of the massive core.  This is 
consistent with the idea that star formation is a highly dynamical
process (Pringle 1989; Elmegreen 2000; Hartmann, 
Ballesteros-Paredes \& Bergin 2001).  Although the calculation
described here does not include magnetic fields, calculations
by other authors including magnetic fields also find that the dissipation
of turbulent support and gravitational collapse occur on the 
dynamical timescale of the clouds they model 
(e.g.\ Ostriker et al.\ 2001; Li et al.\, in preparation).  
Hence, the fact that we have neglected magnetic fields 
does not invalidate this conclusion.

\subsection{Stellar velocity dispersion and distribution}
\label{stellarveldisp}

The main dense core forms a small cluster of stars and brown dwarfs.
This cluster never contains more than $\approx 16$ objects, so
its timescale for dissolution is similar to that on which the
objects are forming.  The objects undergo chaotic interactions 
and eventually disperse with 
stars and brown dwarfs ejected in random directions from the core.
The velocities of the stars and brown dwarfs relative to the 
centre-of-mass velocity of all the objects are given in 
Figure \ref{veldisp}.
The root mean square velocity dispersion is 2.1 km/s in three
dimensions (3-D) or 1.2 km/s in one dimension (1-D) (using the
centre-of-mass velocity for close binaries).  This is roughly
a factor of three greater than the 3-D velocity dispersion 
of the gas when the stars begin to form (${\cal M}=3.8$ giving a 
3-D velocity dispersion of 0.7 km/s).  Thus, dynamical 
interactions are the primary source of the overall
stellar velocity dispersion, but they do not often result in 
extreme ejection velocities.

The velocity dispersion of the stars and brown dwarfs is
similar to that observed in low-mass star-forming regions.  
For example, the 1-D velocity dispersion in Taurus-Auriga has been
measured at $\lsim 2$ km/s using proper motions 
(Jones \& Herbig 1979; Hartmann et al.\ 1991; Frink et al.\ 1997).
The radial velocity dispersion of stars in Chamaeleon I has been 
measured at $\approx 3.6$ km/s \cite{JoeGue2001}, but it is thought 
that this value is high due to the radial velocity `noise' exhibited
by T Tauri stars \cite{Guentheretal2001}.  For the Orion Trapezium 
Cluster, the 1-D velocity dispersion is $\approx 2.3$ km/s 
(Jones \& Walker 1988; Tian et al.\ 1996).  This is somewhat greater than 
in our calculation, but the stellar cluster that we form also
has a much lower mass.

We find the velocity dispersion is independent both of stellar 
mass and of binarity (Figure \ref{veldisp}).  The lack of 
dependence on mass has been found in N-body simulations
of the break up of small-N clusters with $N>3$ \cite{SteDur1998}
and in SPH calculations of N=5 clusters embedded in gas 
(Delgado-Donate, Clarke \& Bate 2003).
Essentially, as $N$ increases, the escape speed is determined by
the gravitational potential well of the group rather than by individual
stellar masses.
However, both Sterzik \& Durisen and Delgado-Donate et al.\ 
found that binaries should have a smaller 
velocity dispersion than single objects due to 
the recoil velocities of binaries being lower.  This is not 
what we find.  Part of this is due to the fact that 
the stellar velocities in our calculation are contributed to by
the motions of the dense cores (i.e.\ turbulent motions in the gas), 
something that was not considered in the above studies
because they modelled only isolated small-N clusters.  
Another factor is that the most massive core in our calculation 
produces a large number of objects ($N=38$) and the presence of
gas allows dissipative interactions between stars.  Thus, the core forms 
several binaries, some of which are ejected,
essentially eliminating any difference in the velocity 
dispersions of binaries and single objects.  The formation of
several binaries is difficult in N-body simulations 
(e.g.\ Sterzik \& Durisen 1998) due to the absence
of dissipative interactions between the stars and gas, 
while in $N=5$ clusters with gas (Delgado-Donate et al.\ 2003) 
it is unlikely simply due to the small number of objects.

Both Sterzik \& Durisen \shortcite{SteDur1999} and Reipurth \& Clarke 
\shortcite{ReiCla2001}
suggested that brown dwarfs might have large radial velocities 
due to their formation in and subsequent ejection from multiple 
systems.  However, although all the brown dwarfs formed in 
our calculation do result from the ejection of low-mass objects 
from unstable multiple systems (see Section \ref{BDform}), our results 
show that this mechanism does not generally produce large ejection
velocities.  This is confirmed observationally by Joergens \& 
Guenther \shortcite{JoeGue2001} who studied the radial velocities 
of brown dwarfs in the Chamaeleon I dark cloud and found a 
velocity dispersion of only 2 km/s.  One potential area of concern is that
we soften the gravitational fields of stars and brown dwarfs on scales
less than 4 AU (Section 3.2).  Thus, ejection velocities from very close
encounters will be underestimated.  However, we have plotted the final
velocities of stars and brown dwarfs versus their closest encounter 
distance and
find no correlation (either for objects with closest encounters greater
than or less than 4 AU).  The reason for this is probably the same
as that which results in the independence of the velocity dispersion on mass,
namely that the most likely ejection velocity from a stellar group 
depends on the escapee having sufficient kinetic energy to escape 
the gravitational potential well of the group rather than the details
of individual interactions.  Furthermore, the escape velocity of two
brown dwarfs separated by 4 AU is $\approx 7$ km/s which is 
significantly higher than the velocity dispersion we find.  Thus,
we can safely conclude that our results are not significantly affected
by our gravitational softening length.

Although we find that the 3-D velocity dispersion is quite
low, the objects are still able to travel $\sim 0.2$ pc (the 
radius of the initial cloud) in $10^5$ years (the timescale over
which star formation occurs).  Therefore, at the end of the 
calculation, the cores are surrounded by a halo of objects.  
An observer looking at the positions of these stars on the 
sky would have no idea whether these stars formed in their 
current locations or in the dense cores.  Only by 
determining their 3-D space velocities to high precision
could our prediction that stars form in dense cores and are 
ejected from them be tested observationally.

Nevertheless, there is some observational support for 
the concept of dense cores resulting in small, expanding 
associations of stars.  
For example, in Taurus the 6 Gomez groups \cite{Gomezetal1993}
have radii of $\approx 0.5-1.0$ pc and contain $\approx 10-20$ stars each.
With 1-D expansion velocities of $\sim 1$ km/s they need only
be $0.5-1$ million years old.
A more recent survey of pre-main-sequence stars in Ophiuchus 
\cite{Allenetal2002} finds that the dense cores are surrounded 
by young stars within one or two tenths of a parsec.  It is 
intriguing to speculate that these stars formed within the observed
dense cores and were ejected to occupy their current locations while 
more stars are currently being formed in the dense cores in
new `bursts' of star formation (as discussed in Section
\ref{efficiencytimescale}).
The stars observed by Allen et al.\ could have travelled to 
their current locations in $\lsim 2\times 10^5$ years.

\begin{figure*}
\centerline{\psfig{figure=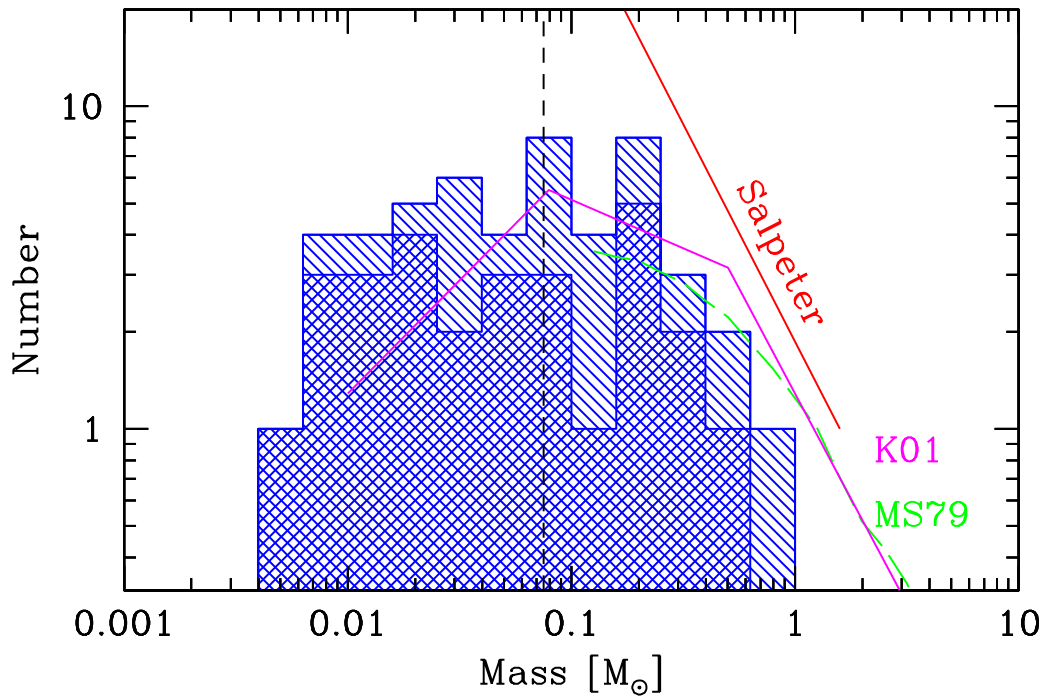,width=15.0truecm,height=10.03truecm,rwidth=15.0truecm,rheight=10.03truecm}}
\caption{\label{imf} The initial mass function (IMF) at the end of the simulation.  The single shaded region shows all of the objects, the double shaded region shows only those objects that have finished accreting.  The mass resolution of the simulation is 0.0011 M$_\odot$ (i.e.\ 1.1 M$_{\rm J}$), but no objects have masses lower than $5$ M$_{\rm J}$ due to the opacity limit for fragmentation.  The resulting IMF is consistent with Salpeter above 0.5  M$_\odot$, and flat below this with a sharp cutoff due to the opacity limit for fragmentation at $\approx 0.006$ M$_\odot$.  We also plot fits to the observed IMF from Miller \& Scalo (1979) (dashed line) and Kroupa (2001) (solid line).  The Salpeter slope is equal to that of Kroupa (2001) for $M>0.5$ M$_\odot$.  The vertical dashed line marks the star/brown dwarf boundary.}
\end{figure*}

\subsection{Initial mass function}
\label{imfsec}

In Figure \ref{imf} we plot the initial mass function (IMF) obtained from the 
calculation.  This is the first initial mass function to be 
determined from a hydrodynamical calculation that resolves objects
down to (and beyond) the opacity limit for fragmentation.  Hence, it predicts
both the stellar and sub-stellar IMF.  It is worth pointing out
that the number of objects formed (50 stars and brown dwarfs)
is larger than the number of objects observed in some surveys of 
star-forming regions.

We obtain a mass function that is consistent with
\begin{equation}
\frac{{\rm d}N}{{\rm d log}M} \propto M^{\Gamma}
\end{equation}
where
\begin{equation}
\Gamma = \left\{ \begin{array}{rl}
                -1.35 &\mbox{\rm for $M\gsim 0.5$~M$_\odot$} \\
                0.0 & \mbox{\rm for $0.006 < M \lsim 0.5$~M$_\odot$}
               \end{array}
         \right.
\end{equation}
and there are no objects below the opacity limit for fragmentation
($\approx 0.005$~M$_\odot$).
The Salpeter slope is $\Gamma=-1.35$ \cite{Salpeter1955}.

In the present calculation, the effect of the opacity limit is mimicked 
by the switch from an isothermal equation of state to $\eta=7/5$ at a
density of $10^{-13}$ g~cm$^{-3}$ (equation \ref{eta}).  
This is only an approximation of the true behaviour of the 
gas.  In reality, the opacity limit depends on the heating and 
cooling rates of collapsing gas (Section 2).  In order to get
the exact value of the opacity limit for fragmentation correct, 
calculations
including full radiative transfer are required.  Thus, while 
the precise mass at which the cutoff in the IMF occurs cannot 
be derived from the present calculation, the qualitative result 
that there is a sharp cutoff in the mass function due to the 
opacity limit at $\sim 0.01$ M$_\odot$ is expected.

\subsubsection{Comparison with the observed IMF}

How does this theoretical IMF compare with the observed IMF?
It is generally agreed that the observed IMF has a Salpeter-type
slope above $\approx 0.5$ M$_\odot$ and a flatter slope below 
this mass (e.g.\ Kroupa 2001; Luhman et al.\ 2000). 
Our theoretical 
IMF is consistent with these observations, although the relatively 
small number of stars and the fact that we don't form any stars 
with masses $\gsim 0.75$ M$_\odot$ means that we cannot say
much about the slope above 0.5 M$_\odot$.  On the other hand, there 
is no a priori reason that the calculation could not have produced,
say, one 10-M$_\odot$ star (or even a 2-M$_\odot$ star) 
and a cluster of brown dwarfs, something that would have been 
inconsistent with the observed IMF. 

The real prediction of this calculation is for the low-mass 
stellar and substellar portions of the IMF where, unlike 
observations, we can detect all the brown dwarfs that form.  
This region of the IMF, particularly the substellar portion, 
is currently attracting an enormous amount of observational 
effort, and its exact form is still open to debate.
Kroupa \shortcite{Kroupa2001} considered many observational
surveys and concluded that the IMF rises slowly from 0.5 M$_\odot$
to the stellar/substellar boundary 
($\Gamma=-0.3$) and decreases with $\Gamma\approx 0.7$
in the brown dwarf regime (Figure \ref{imf}).  The most recent
surveys appear to be converging to a substellar slope of
$\Gamma\approx 0.5$ (Hillenbrand \& Carpenter 2000; 
Najita, Tiede \& Carr 2000; Moraux, Bouvier \& Stauffer 2001).  
With our small number of objects, the best we can say at the 
moment is that our theoretical IMF is roughly flat 
in the range $0.01-0.5$ M$_\odot$ and is consistent with these 
observational results.

A flat slope from $0.01-0.5$ M$_\odot$ implies that there are 
roughly equal numbers of stars and brown dwarfs (Figure \ref{imf}; Table 1).
Observational surveys in the local solar neighbourhood agree
with this prediction; both Reid et al. \shortcite{Reidetal1999} 
and Chabrier \shortcite{Chabrier2002}
estimate that the numbers of stars and brown dwarfs are roughly equal.

In Section 5.1, we likened our calculation to modelling part of 
the $\rho$ Ophiuchus star-forming region.
Several papers have considered the low-mass IMF in $\rho$ Ophiuchus.
Comeron et al.\ \shortcite{Comeronetal1993} and 
Strom, Kepner \& Strom \shortcite{StrKepStr1995} 
both estimated $\Gamma=0$ in the 
range $0.05 - 1$ M$_\odot$ from luminosity 
functions.  Luhman \& Rieke (1999) find a break in the IMF at 
$\approx 0.5$ M$_\odot$ and estimate a lower limit of $-0.5$ for 
the slope in the range $0.02 - 0.5$ M$_\odot$.  Finally, the
HST/NICMOS survey of Allen et al.\ \shortcite{Allenetal2002} 
estimates that 30\% of
the objects in $\rho$ Ophiuchus may be brown dwarfs.  Thus, our
results are consistent with the IMF in Ophiuchus.

\subsubsection{The origin of the IMF}

There is an ongoing discussion regarding the origin of the IMF
(see the review by Kroupa 2002).
Some suggest that it is due to stellar feedback 
(e.g.\ Silk 1977b; Adams \& Fattuzzo 1996),
others that it is due to the density structure in molecular gas 
(e.g.\ Larson 1978; 
Motte et al.\ 1998; Elmegreen 2002, Padoan \& Nordlund 2002),
and yet others attribute it to competitive accretion of gas and dynamical
interactions (e.g.\ Larson 1978; Zinnecker 1982; Bonnell et al. 1997, 2001a,b; 
Klessen et al.\ 1998; Klessen \& Burkert 2000).

The calculation presented here does not include feedback processes,
yet still produces a realistic IMF.  Moreover, there is no direct
mapping between the instantaneous density structure in the gas and 
the masses of the stars that result.  Although the stars form 
within dense cores, their individual masses are determined by 
accretion over the same time-scale on which the gas
distribution in the core evolves.  Furthermore, with each core 
containing multiple stars, there is no one-to-one link between 
core masses and final stellar masses.  Rather, the origin of the 
IMF is best described as resulting from the combination of
cloud structure, competitive accretion, and dynamical encounters
that eject the stars from the dense cores and terminate their
accretion (Bonnell et al.\ 1997, 2001a,b; Klessen et al.\ 1998; 
Klessen \& Burkert 2000, 2001).  It is a highly dynamical, chaotic
process that needs to be understood in a statistical manner.

To determine how the initial mass function depends on environment
will require further large-scale calculations in which the 
initial conditions such as the thermal Jeans mass and the 
density and velocity structure of the gas are varied.  From the current 
calculation, we find that, if we take our `best guess' values 
for the initial conditions in local star-forming molecular 
clouds, we do reproduce the observed IMF.

\subsection{Brown dwarfs}

Even though the initial thermal Jeans mass in the cloud is 
1 M$_\odot$, the calculation produces many brown dwarfs with masses
as low as $0.005$ M$_\odot$.  Even if we take into account the 
initial velocity dispersion of the gas, with Mach number ${\cal M}=6.4$, 
the local Jeans mass in isothermal shocks is still 
$\approx 0.16$ M$_\odot$ 
and by the time the stars and brown dwarfs begin to form the 
Mach number has dropped to ${\cal M}=3.8$.
Therefore, it is of interest to determine how the brown dwarfs form.
In particular, do they form in the same manner as 
the stars or in some other way?

\subsubsection{The formation of brown dwarfs}
\label{BDform}

The formation mechanism and resulting properties of the brown
dwarfs in this calculation have been studied in detail in the 
companion paper, Bate et al.\ \shortcite{BatBonBro2002a}.
We find that all 18 of the definite brown dwarfs (i.e.\ those 
that are not accreting significantly at the end of the calculation) 
form in dynamically-unstable multiple systems and are ejected from 
the regions of dense gas in which they form before 
they can accrete enough gas to become stars.  We emphasise
that {\it all objects begin as opacity-limited fragments containing only a
few Jupiter masses} (Section \ref{sinkparticles}).  
Those that subsequently become stars accrete
large quantities of gas from the dense cores in which they form, while
those that remain as brown dwarfs do not because they are ejected.

Watkins et al.\ \shortcite{Watkinsetal1998} speculated that 
brown dwarfs may be ejected from unstable multiple systems,
but they did not recognise the importance of these ejections
in halting accretion.  Reipurth \& Clarke \shortcite{ReiCla2001}
proposed that brown dwarfs form because dynamical ejections 
halt the accretion onto objects that would otherwise become 
stars.  This is the mechanism by which all our brown dwarfs 
form.  However, without numerical calculations, the main processes 
involved in the formation of these unstable multiple systems, the
efficiency of the ejection mechanism, and the resulting properties of 
the brown dwarfs can only be conjectured.

We find (Bate et al.\ 2002a) that roughly 
three quarters of the brown dwarfs (14 of the 18) form the 
via the fragmentation of gravitationally-unstable circumstellar discs 
(Bonnell 1994; Whitworth et al.\ 1995; Burkert et al.\ 1997).
The remaining brown dwarfs (4 of the 18) form via the fragmentation 
of filaments of molecular gas (e.g.\ Bonnell et al.\ 1991).
These objects either form in, or quickly fall into, unstable 
multiple systems are subsequently ejected from the dense gas, 
limiting their masses to be substellar.

Examining the origins of the 23 stars formed in the calculation,
we find that only one third (7 of the 23) form via disc fragmentation.
The majority (16 of the 23) form directly from the collapse of the
cloud in filaments of molecular gas.  Thus, both stars
and brown dwarfs can form by both mechanisms, but the primary
mechanism by which stars form is the fragmentation of collapsing 
gas while the primary mechanism for brown dwarf formation is 
disc fragmentation.
As proposed by Reipurth \& Clarke \shortcite{ReiCla2001}, 
the main difference between 
stars and brown dwarfs is that the
brown dwarfs are ejected soon after they are formed, before
they have been able to accrete to stellar masses.  This is more
likely if a disc fragments into multiple objects than
if an object forms on its own and has to collide with a nearby 
multiple system before it is ejected from the dense gas.  Thus,
the different fractions of stars and brown dwarfs that form
from cloud and disc fragmentation are explained.

The ease with which disc 
fragmentation occurs depends primarily on the rate at which 
it accretes mass from the surrounding cloud (Bonnell 1994) 
and the disc's equation of state (e.g.\ Pickett et al.\ 2000).  
The density of the gas in the discs that fragment to form brown
dwarfs is high enough that the gas is in the $\eta=7/5$ regime
(Section 2.1).  Thus, the gas resists fragmentation far more than 
it would if an isothermal equation of state were used, although 
our equation of state does not include heating from shocks.  
On the other hand, because the flattened disc geometry may allow 
more rapid cooling than a spherically-symmetric geometry, real 
discs may be cooler and more unstable than those 
we model here.  Hence, the number of brown dwarfs may be even 
greater.  Another factor is that stars and brown dwarfs cannot 
merge in our calculation.  Mergers may reduce the final number of
brown dwarfs.  In summary, a more definitive 
prediction will have to wait until a large-scale calculation 
is performed with radiative transfer and even higher resolution.
For the present, we have demonstrated that the 
fragmentation of a turbulent molecular cloud is capable of 
forming similar numbers of brown dwarfs and stars as well as 
an IMF that is in agreement with observations (Section \ref{imfsec}).

\subsubsection{The frequency of binary brown dwarfs}
\label{BDbin}

\begin{table*}
\begin{tabular}{lcccccl}\hline
Object Numbers & $M_1$ & $M_2$  & $q$ & $a$  & $e$  & Comments \\
        & M$_\odot$ & M$_\odot$ &  & &  \\ \hline
3,10    & 0.73  & 0.41  & 0.56  & 1.1*   & 0.68*  & \\ 
7,8     & 0.54  & 0.24  & 0.44  & 2.0*   & 0.94*  & System 1; Ejected from cloud\\ 
20,22   & 0.35  & 0.11  & 0.33  & 2.2*   & 0.87*  & \\ 
44,42   & 0.10  & 0.095 & 0.93  & 2.6*   & 0.99*  & \\
26,40   & 0.13  & 0.039 & 0.29  & 6.7*   & 0.97*  & Star/brown dwarf binary\\
39,41   & 0.070 & 0.047 & 0.67  & 5.7*   & 0.72*  & Binary brown dwarf\\
45,38   & 0.083 & 0.079 & 0.96  & 8.8*   & 0.59*  & \\ \hline
(3,10),35 & (1.14) & 0.083 & 0.72 & 28    & 0.67  & \\
(20,22),25 & (0.46) & 0.23 & 0.50 & 28    & 0.45  &  \\
32,(44,42)& 0.24 &(0.20)& 0.83  & 30    & 0.02  & \\
(45,38),43& (0.16)&0.022 & 0.14 & 90    & 0.90  & \\ \hline
(26,40),(39,41) &(0.17)&(0.12)&0.68&84  & 0.45  &  \\
(32,(44,42)),50 &(0.44)&0.008 &0.02&54  & 0.31  & System 2; In core 3\\
((3,10),35),46  &(1.23)&0.031 &0.03&354 & 0.44  & \\
((20,22),25),24 &(0.69)&0.17  &0.25&257 & 0.85  &  \\\hline
(((3,10),35),46),23 & (1.26)&0.032&0.03&348 &0.70 & \\
(((20,22),25),24),37& (0.86)&0.022&0.03&226 &0.28 & \\\hline
((((20,22),25),24),37),27& (0.89)&0.005&0.006&643 &0.33 & \\\hline
((26,40),(39,41)),((45,38),43) & (0.29) & (0.18) & 0.64 & 328 & 0.42 & System 3; In core 2 \\\hline
((((3,10),35),46),23),(((((20,22),25),24),37),27) & (1.29)&(0.89)& 0.69& 666 & 0.30 & System 4; In core 1 \\
\hline
\end{tabular}
\caption{\label{table2} The properties of the 4 multiple systems with semi-major axes less than 1000 AU formed in the calculation (see also Figures 12 and 13).  These systems have 2, 4, 7, and 11 members.  The structure of each system is described using a binary hierarchy.  For each `binary' we give the masses of the primary $M_1$ and secondary $M_2$, the mass ratio $q=M_2/M_1$, the semi-major axis $a$, and the eccentricity $e$.  The combined masses of multiple systems are given in parentheses.  Orbital quantities marked with asterisks are unreliable because these close binaries have periastron distances less than the gravitational softening length.  When the calculation is stopped, the three high-order systems are unstable and/or are still accreting, so their final states are unknown.  }
\end{table*}

We find a low frequency ($\sim 5$\%) of binary brown dwarf systems
(Bate et al.\ 2002a).  This is primarily due to the closeness of
the dynamical encounters that eject the brown dwarfs from the 
dense gas in which they form before they can accrete to stellar 
masses.  The minimum separations during the encounters
are usually less than 20 AU (Section 
\ref{discs}), so any wide systems are usually disrupted.  However,
another type of dynamical interaction also plays a role.
Several binary brown dwarf systems that form during the calculation
are destroyed by exchange interactions where one or both of the
brown dwarfs are replaced by stars.  

These two effects result in the low frequency of binary brown dwarfs.
Of the 18 definite brown dwarfs, none are in binaries.  However,
at the end of the calculation there is a close binary brown 
dwarf (semimajor axis 6 AU) within an unstable multiple system
(Figure \ref{core2}, $t=1.40 t_{\rm ff}$; Table \ref{table2}, 
binary 39,41).  
This system will undergo further dynamical evolution,
and it is still accreting.
Since the binary brown dwarf is very close, it is possible that 
it will survive the dissolution of the multiple system and 
be ejected before it has become a stellar binary.  Even so, this 
would result in only one binary brown dwarf system and 
$\approx 20$ single brown dwarfs.  Thus, the formation of close 
binary brown dwarfs is possible, but the fraction of 
brown dwarfs with a brown dwarf companion should be low ($\sim 5$\%).

Observationally, the frequency of brown dwarf binaries is not 
yet clear.  Both Reid et al.\ (2001) and Close et al.\ (2002)
observed 20 brown dwarf primaries and found that 4 have companions 
giving binary frequencies of $\approx 20$\%.  However, as 
discussed by Close et al., these surveys are magnitude limited 
rather than volume limited and therefore are likely to overestimate
the true frequency of brown dwarf binaries.  On the other hand, 
these surveys cannot detect faint companions or those at very
small separations, which would boost the frequency.  Our calculation
favours a lower frequency, but due to our small number of objects, 
we cannot exclude a frequency of 20\% (there would be a probability
of $\approx 6$\% of finding 1 binary out of 20 systems).
It is important
to note that none of the binary brown dwarf systems currently 
known have projected separations $> 15$ AU 
(Reid et al.\ 2001; Close et al.\ 2002).  
This is consistent with their having survived dynamical ejection 
from unstable multiple systems.

\begin{figure*}
\centerline{\psfig{figure=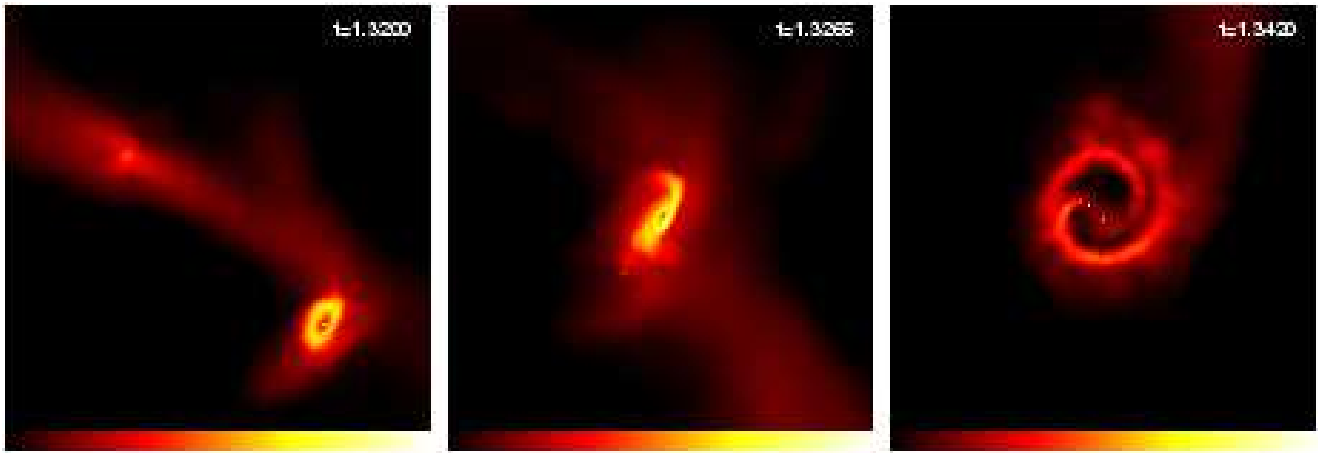,width=17.5truecm,height=6.02truecm,rwidth=17.5truecm,rheight=6.1truecm}}
\caption{\label{stardisc} A star-disc interaction resulting in the formation of a triple system.  Two fragments form separately (left).  They fall together and undergo a star-disc encounter at $t\approx 1.325 t_{\rm ff}$.  This encounter produces an eccentric binary system.  During the second periastron passage the interaction results in the fragmentation of the disc to form a third object.  This system evolves chaotically and forms a triple system surrounded by a circumtriple disc ($t=1.3420 t_{\rm ff}$).  Each panel is 400 AU across.  Time is given in units of the initial free-fall time of $1.90\times 10^5$ years.  The panels show the logarithm of column density, $N$, through the cloud, with the scale covering $0.5 < \log N < 3.5$ with $N$ measured in g cm$^{-2}$.}
\end{figure*}

\subsubsection{The sizes of discs around brown dwarfs}
\label{BDdisc}

Another result of the close dynamical encounters that occur
during the ejection of the brown dwarfs from the dense gas is that
most brown dwarfs do not have large discs.
We find that only one ($\sim 5$\%) of the definite brown dwarfs is
ejected with a large (radius $\gsim 20$ AU) circumstellar disc.
Of the 18 definite brown dwarfs, 14 have dynamical 
encounters at separations $<23$ AU (Section \ref{discs}).
However, this is not the only reason that few brown dwarfs have 
large discs.  To become brown dwarfs they must, by definition, be
ejected from the dense gas before they have been able to accrete
much gas.  Three of the definite brown dwarfs are ejected so 
soon after their formation that they have not had time to accrete 
gas with the high specific angular momentum required to form 
large discs.  Only one of the 18 definite brown dwarfs 
is ejected with a resolved disc 
(radius $\approx 60$ AU).  The other brown dwarfs will possess 
smaller discs, but in this calculation we are unable 
to resolve discs with radii $\lsim 10$ AU (Section 3.2).  
In any case, truncation radii $\ll 10$ AU are not 
observationally relevant since the viscous timescale at this radius is
\begin{equation}
\begin{array}{ll}
t_{\rm visc} & \displaystyle \sim \frac{1}{\alpha} \left(\frac{R}{H}\right)^2 2\pi \sqrt{\frac{R^3}{GM}} \\ \\ & \displaystyle = 1 \times 10^6 \left( \frac{R}{10 {\rm AU}}\right)^{3/2} \left( \frac{0.075~{\rm M}_\odot}{M}\right)^{1/2} {\rm yr}
\end{array}
\end{equation}
where we have assumed the standard Shakura-Sunyaev (1973) 
viscosity parameter $\alpha = 0.01$ (Hartmann et al.\ 1998) 
and the disc height to radius $H/R = 0.1$ 
(e.g.\ Burrows et al.\ 1996; Stapelfeldt et al.\ 1998).  
Thus, any disc that is
truncated to a radius $\ll 10$ AU around a brown dwarf
is likely to evolve viscously to a radius of $\approx 10$ AU 
before it is observed.  Of course, such a disc may have a very low
mass, first because of the truncation and second because much of
the remaining gas would be accreted during
its viscous evolution back out to $\approx 10$ AU radius.  
Armitage \& Clarke \shortcite{ArmCla1997} studied the evolution and
observability of discs truncated to between 1 and 10 AU.  They found
that the K-band excess becomes too low to detect within a few times
$10^5$ years, but that infrared emission at wavelengths $\gsim 5$ $\mu$m
is detectable for much longer.

Muench et al.\ \shortcite{Muenchetal2001} find $\approx 65$\% of 
brown dwarfs in the Orion Trapezium cluster have infrared excesses
indicative of discs.  Spectral energy distributions indicative 
of discs have also been found around young brown dwarfs in low
mass star-forming regions (Natta \& Testi 2001; Testi et al.\ 2002) 
as have H$\alpha$ and Br$\gamma$
accretion signatures (Martin et al.\ 2001a; 
Zapatero Osorio et al.\ 2002b).
Our results show that if brown dwarfs are formed by the ejection 
mechanism, few of these discs should have radii $> 20$ AU and 
most discs should have radii $\approx 10$ AU, although those 
that were severely truncated and evolved back out to this 
size might be difficult to detect.

\subsubsection{Brown dwarf companions to stars}
\label{BDstar}

As we mention in Section \ref{evolution}, when the calculation is 
stopped, 9 objects have substellar masses but are still accreting.
All of these potential brown dwarfs are members of unstable
multiple systems (Table 2).  Along with the one close binary brown
dwarf system mentioned above, there is one close stellar/substellar 
binary (26,40 in Table 2).  This system is part of the same unstable 
septuple system
that contains the binary brown dwarf system.  It has a semi-major
axis of 7 AU.  It is known that brown dwarf companions to FGK stars 
with semi-major axes $\lsim 4$ AU are very rare ($\lsim 1$\%; 
Halbwachs et al.\ 2000; Zucker \& Mazeh 2001; 
Pourbaix \& Arenou 2001).
This is the so-called brown dwarf desert.  Our stellar/substellar
system is of interest because it indicates that there may not be
a brown dwarf desert for M stars.  Of course, it has several 
hurdles to pass before it becomes a stable stellar/substellar
system.  It must avoid the accretion of much more gas, which would
tend to equalise the masses \cite{Bate2000}; it must survive the 
break up of the septuple system; and the brown dwarf must avoid
migrating into the star due to its interaction with the disc
\cite{ArmBon2002}.  We note that the
ejection hypothesis for the formation of brown dwarfs is not
necessarily inconsistent with the detection of close brown dwarf
companions to stars.  The hypothesis relies on the assertion 
that brown dwarfs are ejected from the regions of dense gas 
in which they form before they are able to accrete
to stellar masses.  However, there is no reason why they cannot
be in a close orbit around a star when the {\it system} is ejected.
More difficult would be the formation of wide ($\gsim 100$ AU) 
brown dwarf companions to stars by this mechanism since it is 
difficult to see how such a
wide system could be ejected dynamically.  Such systems are
known to exist (Nakajima et al.\ 1995; Rebolo et al.\ 1998;
Lowrance et al.\ 2000; Gizis et al.\ 2001; Els et al.\ 2001).  
In these cases, it may simply be 
that the specific angular momentum of the gas accreted by the 
system was low enough to be preferentially accreted by the
primary, leaving the secondary with a substellar mass
(Bate \& Bonnell 1997; Bate 2000).

\subsection{The formation of binary and multiple systems}

The favoured mechanism for binary star formation is
fragmentation, either of collapsing molecular cloud cores
(Boss \& Bodenheimer 1979; Boss 1986; Bonnell et al.\ 1991;
Nelson \& Papaloizou 1993; Burkert \& Bodenheimer 1993; Bate,
Bonnell \& Price 1995; Truelove et al.\ 1998)
or of massive circumstellar discs 
(Bonnell 1994; Whitworth et al.\ 1995; Burkert et al.\ 1997).  
However, it has been
pointed out that, especially in small groups of stars, star-disc
encounters may form binaries (Larson 1990; Clarke \& Pringle 1991a,b;
Heller 1995; McDonald \& Clarke 1995).  In a star-disc encounter
resulting in capture,
one star passes through the disc of another, dissipating 
enough kinetic energy to form a bound system \cite{HalClaPri1996}.
The calculation presented here can be used to examine which of
these formation mechanisms is most prevalent.

\begin{figure}
\centerline{\psfig{figure=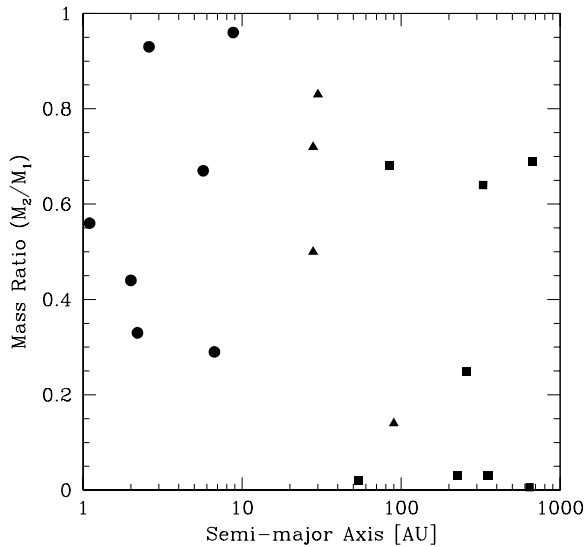,width=8.0truecm,height=8.0truecm,rwidth=8.0truecm,rheight=7.5truecm}}
\caption{\label{closebinary} Mass ratios versus semi-major axes of the multiple systems that exist at the end of the calculation (Table 2).  Binaries are plotted with circles, triples with triangles and higher-order systems with squares.  Note that there is a preference for close binaries to have nearly equal masses, and the only extreme mass ratio systems are wide components of multiple systems still evolving when the calculation was stopped. }
\end{figure}

We find that the dominant mechanism for the formation of the
binary and multiple systems in the calculation is fragmentation, 
either of gaseous filaments or of massive circumstellar discs.  
Although star-disc encounters do occur 
during the calculation, most of these
serve only to truncate the circumstellar discs (see Section \ref{discs}) 
and do not result
in bound stellar systems (c.f.\ Clarke \& Pringle 1991a).  
There are a few exceptions.  One star-disc encounter between a 
binary and a single star results in the formation
of an unstable triple system which subsequently breaks up into a
binary and a single star.  Another star-disc capture is 
depicted in Figure \ref{stardisc}.  It is also important to note that,
although star-disc encounters do not usually form simple 
bound systems directly, they do result in dissipation, which 
aids in the formation of the small-N bound groups that 
later dissolve and produce binary and multiple systems.  Thus,
dissipative encounters do play an important role in the star
formation process, if not in the simple picture of star-disc
capture.

\subsubsection{Multiplicity}

At the end of the calculation, there exist 4 multiple systems 
or stellar groups.  Their properties are displayed in Table 2
and Figure \ref{closebinary}.  Two of these systems originate in
the first dense core.  They are a close binary system that
was ejected from the cloud (7,8) and the remains of a small-N group 
consisting of 11 objects (see also the last panel of
Figure \ref{core1b}).   Each of the other two dense cores
contains a multiple system, one an unstable quadruple system
and the other an unstable system consisting of seven objects
(see the last panels of Figures \ref{core2} and \ref{core3}).

All these systems, except the ejected close binary, will 
undergo further evolution.  It is likely that most of the
close binary systems and some of the triple systems will
survive, but it is not possible to 
determine the final binary and multiple frequencies or 
properties of the wide systems that would eventually 
be formed from the simulation if it were continued.  The 
best we can do is provide an upper limit on the final 
companion star frequency
\begin{equation}
CSF = \frac{B+2T+3Q+....}{S+B+T+Q+....}
\end{equation}
where $S$ is the number of single stars, $B$ is the number of 
binaries, $T$ is the number of triples, etc.  We have
26 single objects,
1 binary, 1 quadruple, 1 septuple and 1 system of 11 objects.
This gives a companion star frequency of $20/30=67$\%.
Alternately, the number of companions divided by the total
number of objects is $20/50=40$\%.  These high frequencies
are in broad agreement with the large fractions of binary
and multiple systems found in young star-forming regions
(Ghez, Neugebauer \& Matthews 1993; Leinert et al.\ 1993;
Richichi et al.\ 1994; Simon et al.\ 1995; Ghez et al.\ 1997;
Duch\^ene 1999).
One potential difficulty with the calculation is that 
there are no wide binary systems when it is stopped.  The multiple
systems are expected to evolve into stable configurations
that are most likely to be binaries or triples.  However,
it is not clear that these systems will be 
wide.  Furthermore, although there are many low-mass
objects in the multiple systems, these are the most likely
components to be ejected in subsequent dynamical interactions.
Thus, any wide binaries that do form may not have low-mass secondaries, 
whereas we know from observations that most wide binaries
do have unequal mass components \cite{DuqMay1991}.  
Similar results are obtained by Delgado-Donate et al.\ (2003) in their
simulations of $N=5$ clusters embedded in molecular cloud cores.
These issues need to be addressed in future calculations.

\subsubsection{The formation of close binary systems}

As reviewed in section 2, the opacity limit for fragmentation
sets a minimum initial binary separation of $\approx 10$ AU.
However, at the end of the calculation, there exist 7 close
binary systems (separations $< 10$ AU).  The mechanisms 
by which these close binaries form and their properties have been
discussed in detail in the second companion paper to this, 
Bate et al.\ (2002b).  We only briefly summarise the 
conclusions here.

\begin{figure}
\centerline{\psfig{figure=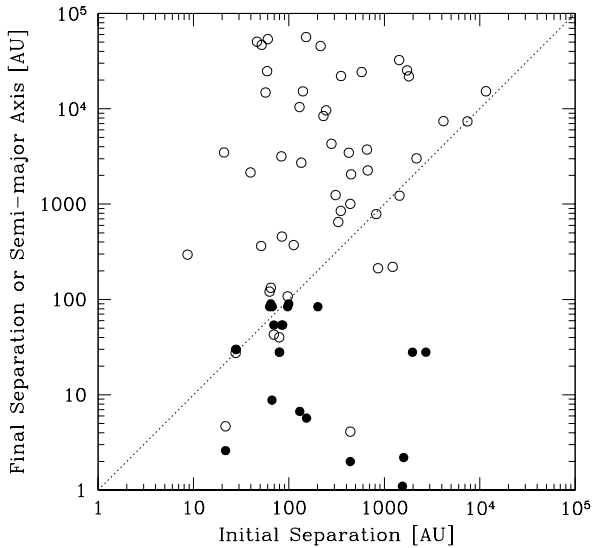,width=8.0truecm,height=8.0truecm,rwidth=8.0truecm,rheight=7.5truecm}}
\caption{\label{separations} We find the closest object to each star or brown dwarf when it forms and plot their final versus initial separation (open circles).  We also plot the final semi-major axes versus the initial separations of all objects that have orbits with semi-major axes less than 100 AU (see Table 2) at the end of the calculation (filled circles).  Note that the closest object when a star or brown dwarf forms does not usually remain close.  Also, most of the close multiple systems at the end of the calculation are composed of objects that formed at large distances from each other.  These results indicate the importance of dynamical interactions during the calculation.}
\end{figure}

Bate et al.\ (2002b) analyse the fragmentation that occurs in
the hydrodynamical calculation and find that the
three smallest separations between an existing object and a 
forming fragment are 9, 21, and 22 AU (Figure \ref{separations},
open circles).  These separations are consistent with the 
lower limit set by the opacity limit for fragmentation.
Only the last of these ends up in a close binary 
system; the other systems are
disrupted by dynamical encounters.  Bate et al.\ (2002b) find that,
rather than forming directly via fragmentation,
the close binary systems form
through a combination of accretion, the interaction of binaries
and triples with circumbinary and circumtriple discs, and 
dynamical interactions.  Accretion onto a binary from an 
envelope decreases 
the binary's separation unless the specific angular momentum
of the accreted material is significantly greater than that of the binary
(Artymowicz 1983; Bate 1997; Bate \& Bonnell 1997; Bate 2000).
It can also change the relative separations of a triple system, 
destabilising
it and forcing dynamical interactions (Smith, Bonnell \& Bate 1997).
Circumbinary discs can remove large amounts of orbital angular 
momentum from an embedded binary system via gravitational torques,
thus tightening its orbit (Pringle 1991; Artymowicz et al.\ 1991).
However, although both of these processes play a role, the most 
important ingredient for the formation of the close binaries
in our calculation is stellar dynamical interaction 
(Bate et al.\ 2002b).  Dynamical interactions can
harden existing wide binaries by removing angular momentum and energy
from their orbits.  They also produce exchange interactions in which
a temporary unstable multiple system decays by ejecting one of the
components of the original binary.  Usually the lowest-mass object is
ejected.  Although such dynamical interactions play the dominant role
in forming the close binaries, they cannot produce the observed 
close binary frequency on their own.  Kroupa \& Burkert 
\shortcite{KroBur2001} find that N-body calculations which begin 
with star clusters (100 to 1000 stars) consisting entirely of 
binaries with periods $4.5 < \log{(P/{\rm days})} < 5.5$ produce 
almost no binaries with periods $\log{(P/{\rm days})}<4$.  Similarly, the
dissolution of small-N clusters typically results in binaries with 
separations only an order of magnitude smaller than the size of the initial
cluster (Sterzik \& Durisen 1998).  The main difference between these
calculations and ours is that the dynamical interactions in our calculation
are usually dissipative.  Along with the effects of accretion and the
interaction of multiple systems with circumstellar discs described above,
the presence of gas allows dynamical encounters to be dissipative and 
to transport angular momentum.
Such dissipative encounters include star-disc encounters 
(Larson 1990; Clarke \& Pringle 1991a,b; McDonald \& Clarke 1995; 
Heller 1995; Hall, Clarke \& Pringle 1996) 
and other tidal interactions (Larson 2002).

These processes of accretion, interaction with circumbinary and 
circumtriple discs and dissipative dynamical interactions result 
in the formation
of seven close binaries by the time the calculation is stopped.  
If all of these survive the break up of the remaining multiple 
systems, the resulting frequency of close binaries would be 
$7/43 \approx 16$\%.  This is in good agreement with the observed 
frequency of close (separation $<10$ AU) binaries of 
$\approx 20$\% \cite{DuqMay1991}, demonstrating
that close binaries need not be created by fragmentation in situ.

\subsubsection{The properties of close binary systems}

The formation mechanisms discussed above lead to several consequences for the
properties of close binaries (Bate et al.\ 2002b).  
We find a preference for equal masses
(Table 2, Figure \ref{closebinary}), with all
close binaries having mass ratios $q \gsim 0.3$ and most having $q>1/2$.
This is due to the mass-equalising effect of the accretion of gas with
high specific angular momentum (Artymowicz 1983; Bonnell
\& Bastien 1992; Whitworth et al.\ 1995; Bate \& Bonnell 1997; 
Bate 2000)
and dynamical exchange interactions that usually result in the ejection
of the least massive component.  These processes give a natural
explanation for the observation that close binaries (periods $\lsim 10$
years) tend to have higher mass ratios than wider binaries 
(Mazeh et al.\ 1992; Halbwachs, Mayor \& Udry 1998; Tokovinin 2000).
The wider multiple systems in the calculation have a larger range
of mass ratios (Figure \ref{closebinary}), although it is 
unclear how this would change with further evolution.

Successive dynamical exchanges also lead to a dependence of the close 
binary fraction on primary mass, since each time a binary encounters 
a star more massive than the primary, the most massive star will usually
become the new primary.  Of the $\approx 20$ brown dwarfs there
is only one close binary brown dwarf system (Section \ref{BDbin}), whereas 
5 of the 11 stars with masses $>0.2$ M$_\odot$ are members of close
binary systems.  While it is difficult to extrapolate these results
to larger star clusters and more massive stars,
this trend of the frequency of close 
binaries increasing with stellar mass is supported by observational surveys 
(Garmany et al.\ 1980;
Abt et al.\ 1990; Morrell \& Levato 1991; Mason et al.\ 1998).

\begin{figure*}
\centerline{\psfig{figure=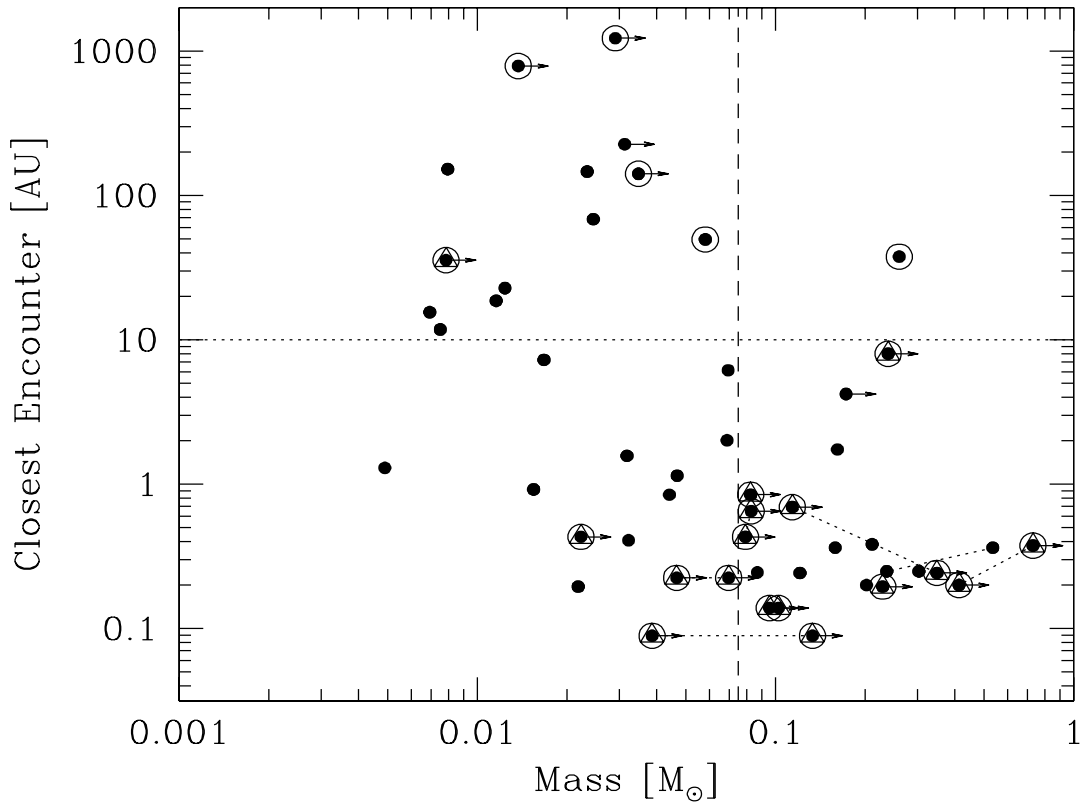,width=15.0truecm,height=12.39truecm,rwidth=15.0truecm,rheight=11.5truecm}}
\caption{\label{encounter} The closest encounter distance of each 
star or brown dwarf during the simulation versus 
its final mass.  Objects that 
are still accreting significantly at the end of the calculation
are denoted with arrows indicating that they are still evolving and
that their masses are lower limits.  Objects that
have resolved discs at the end of the simulation are circled.
Discs smaller than $\approx 10$ AU (horizontal dotted line) cannot 
be resolved by the simulation.  Objects that have had close 
encounters may still have resolved discs due to subsequent accretion
from the cloud.  Note that there are only 11 resolved discs 
at the end of the simulation, but many surround binary and 
higher-order multiple systems (hence the 23 circles in the figure).
Close binaries (semi-major axes 
$<10$ AU) are plotted with the two components connected by dotted 
lines.  Components of multiple systems whose orbits have semi-major 
axes $10<a<100$ AU are denoted by triangles.  All but one of the 
close binaries has a resolved circumbinary or circumtriple disc.  
Encounter distances less than 4 AU are upper limits since the point 
mass potential is softened within this radius.  The vertical 
dashed line marks the star/brown dwarf boundary.  The five
brown dwarfs in the top left corner of the
figure that are still accreting were the last five objects to form
before the calculation was stopped and are thus still evolving 
rapidly.  They may not end up as brown dwarfs or with resolved
discs.  There are 18 brown dwarfs that have finished accreting, 
only one of which has a resolved disc. }
\end{figure*}

At the end of the calculation, only one of the close binaries has 
been ejected from the cloud and is on its own.
The others remain members of larger-scale bound groups
and three are members of hierarchical triple systems (Table 2).
This large number of wider companions
is yet another indication of the importance of multiple systems in 
producing close binaries.  
Even allowing for the eventual dissolution of the bound groups, 
it seems likely that some of the hierarchical triple systems 
will survive.  Although the true frequency of wide companions
to close binaries is not yet well known, many
close binaries do have wider components (e.g.\ Mayor \& Mazeh 1987; 
Tokovinin 1997, 2000).
Indeed, it was this observation that led Tokovinin (1997) to propose that 
dynamical interactions in multiple systems may play an important 
role in the formation of close binary systems.  Further surveys to 
determine the true frequency of wide companions to close binary 
systems would be invaluable.

\subsection{Protoplanetary discs}
\label{discs}

\subsubsection{The formation and evolution of discs}

The calculation resolves gaseous discs with radii greater than
$\approx 10$ AU around the young stars 
and brown dwarfs (see Section 3.2).  We find that 
discs with typical radii of $\sim 50$ AU form around many
of the stars and brown dwarfs soon after their creation 
due to the infall of gas with high specific angular momentum.
Early on, these discs contain a large fraction of the mass
of the star/disc system and are gravitationally unstable
\cite{LinPri1990}.
Spiral density waves form and lead to the rapid transport of
angular momentum outwards and mass inwards via gravitational 
torques (e.g.\ Figures \ref{core2} and \ref{core3}; Bate et al.\ 2002b).  
This serves to increase the mass of the central
object(s) much faster than for a low-mass disc whose evolution
is expected to be driven by the magneto-rotational instability
(Balbus \& Hawley 1991; Hawley \& Balbus 1991).  
Even so, in many cases, the angular momentum
transport via gravitational torques is insufficient to cope
with the rapid infall of gas onto the disc \cite{Bonnell1994}
and the disc fragments to form additional stars and brown dwarfs
(Figures \ref{core2} and \ref{core3}; Sections 5.4 and 5.5; 
Bate et al.\ 2002a).

\begin{table*}
\begin{tabular}{lcccc}\hline
Minimum Encounter Distance & \multicolumn{2}{c}{Number of Objects} & \multicolumn{2}{c}{Resolved Discs} \\\hline
             & Stars & Brown Dwarfs & Stars & Brown Dwarfs  \\\hline
$<1$ AU          & 19 & 4  & 11 & 0\\ 
1-10 AU          & 3  & 6  & 1 & 0\\ 
10-100 AU        & 1  & 6  & 1 & 1 \\ 
100-1000 AU      & 0  & 2  & 0 & 0 \\ \hline
\end{tabular}
\caption{\label{tableenc} The numbers of stars (total) and definite brown dwarfs (those ejected from the cloud) that have had encounters in the given separation ranges by the end of the calculation.  Also, the numbers of these objects that have resolved discs (radii $\gsim 10$ AU) at the end of the calculation.  For example, six of the definite brown dwarfs had encounters with minimum separations in the range $10-100$ AU and only one of these is ejected with a resolved disc. Note that some of the stars are components of binaries with circumbinary discs.  In these cases, each star is counted as having a disc so that the numbers of discs in the table add up to more than the number of resolved discs.}
\end{table*}

\begin{table*}
\begin{tabular}{lll}\hline
Disc Radius & Encircled Objects & Comments \\
\multicolumn{1}{c}{AU}        &         &          \\\hline
200         & (3,10),35 & Circumtriple disc (Figure 4, $t=1.40 t_{\rm ff}$) \\ 
140         &(32,(44,42)),50    & Circumquadruple disc (Figure 6) \\ 
120         &48       & Substellar object formed near end of calculation, would probably become a star \\ 
100         &(20,22),25    & Circumtriple disc (Figure 4, $t=1.40 t_{\rm ff}$)\\ 
 80         &(45,38),43    & Circumtriple disc (Figure 5)\\ 
 60         &29       & Ejected definite brown dwarf (Figure 4, $t=1.38 t_{\rm ff}$, centre, lower-left)\\ 
 60         &47       & Probable brown dwarf (Figure 4, $t=1.40 t_{\rm ff}$, centre-extreme right)\\ 
 50         &11       & Ejected star (Figure 3, $t=1.16-1.20 t_{\rm ff}$, upper left)\\ 
 40         &26,40    & Circumbinary disc (Figure 5, $t=1.40 t_{\rm ff}$, bottom)\\ 
 30         &39,41    & Circumbinary disc (Figure 5, $t=1.40 t_{\rm ff}$, centre)\\ 
 20         &49       & Substellar object formed near end of calculation (Figure 6, $t=1.40 t_{\rm ff}$, in filament)\\ \hline
\end{tabular}
\caption{\label{tablediscs} The discs around objects that are ejected during the calculation or that exist around objects when the calculation is stopped. Discs with radii $\lsim 10$ AU are not resolved.}
\end{table*}

Although many objects form with large discs initially, 
star-disc encounters truncate the majority of these discs.
In Figure \ref{encounter}, we plot the closest encounter
distance for each object during the calculation 
as a function of its final mass.  Most of the stars and 
brown dwarfs have had encounters closer than 4 AU (recall
that sink particle interactions are softened within 4 AU
so minimum separations less than this are unreliable).
During such an encounter, any large disc is truncated so
that its radius is $\approx 1/3$ of the minimum 
separation during the encounter \cite{Hall1997}.
However, even if an object has a close encounter, it may 
still end up with a large disc.  
New discs can form from gas accreted from the cloud after 
the encounter if the object has not been 
ejected from the cloud.  In some cases, objects repeatedly 
have their discs stripped away and then replenished by new infall.
Evidence for replenished discs can be seen 
in Figure \ref{encounter}, where objects are circled if
they have resolved discs at the time of their ejection 
from the cloud, or at the end of the calculation.  Many
of these objects have had encounters closer than the resolution 
limit for discs.

The formation, truncation and replenishment of discs makes 
the issue of whether a star or brown dwarf ultimately has
a resolved disc very complicated.  In Table \ref{tableenc}, 
we consider those 18 brown dwarfs that have stopped accreting
by the end of the calculation and all of the stars (regardless 
of whether or not they are still evolving, since most stars are 
still evolving when the calculation is stopped).  We find 
that the median closest encounter distance for a brown dwarf 
is $\approx 10$ AU and only 1 of the 18 brown dwarfs is
ejected with a resolved disc (radius $\approx 60$ AU).
The other brown dwarfs will possess smaller 
discs, but in this calculation we are unable 
to resolve discs with radii $\lsim 10$ AU (Section 2.2).
By contrast, most of the stars have had encounters closer than
1 AU, but more than half of these are 
surrounded by resolved discs at the end of the calculation
(many of these are discs around binary and multiple systems).

The different results for the stars and brown dwarfs are due
to the fact that the brown dwarfs must be ejected from the 
cloud before they accrete enough material to become stars.
Three effects are involved.  First, for the brown dwarf to
be ejected before it becomes a star, it must be ejected soon
after it forms.  Three of the 18 definite brown dwarfs are
ejected so quickly that they do not have time to accrete the
high angular momentum gas required to form a large disc.
Second, the encounters that result in ejection are usually
close.  Of the 18 definite brown dwarfs, 14 have undergone dynamical 
encounters with separations less than 23 AU, destroying any 
previously resolved discs.  Finally, by definition, the last 
encounter a brown dwarf has is the one that ejects it
from the cloud.  After this, the disc cannot be replenished
by new accretion from the cloud.

The stars, on the other hand, can remain in the dense gas and
accrete for much longer.  During this time, they may undergo close
encounters and have their discs stripped away but if they remain in
the cloud there is plenty of time for them to accumulate new
discs.  Thus, as seen in Table \ref{tableenc}, the stars 
tend to have even closer encounters than the brown dwarfs
(because they tend to remain in stellar groups longer), but
most still have resolved discs when the calculation is stopped.

In summary, the issue of disc formation is complex.  Brown dwarfs
must quickly be ejected from the cloud, in order to remain as
brown dwarfs.  Thus, they tend to have their original discs, which are 
often severely truncated during the ejection process.  Stars, by definition,
are those objects that remain in the cloud long enough to 
accrete sufficient gas.  Therefore, they have more time in which to
undergo close
encounters.  They typically have smaller minimum encounter distances,
but they also have plenty of time to replenish their discs and 
to accumulate large discs.  The chaotic nature of disc formation,
truncation, and replenishment naturally leads to a broad range
of disc radii and masses.  Thus, it should not be surprising
that the discs around some objects in young clusters are undetectable or
have already dispersed, while other objects with the same apparent
age still have discs (e.g.\ Strom et al.\ 1989; Beckwith et al.\ 1990;
Hillenbrand et al.\ 1998; Haisch et al.\ 2001a,b; Rebull et al.\ 2002;
Armitage, Clarke \& Palla 2002).

\subsubsection{Disc frequencies and sizes}

At the end of the calculation, there are 11 resolved discs.
Their radii and the objects they encircle are listed 
in Table \ref{tablediscs}.
Six of the discs surround multiple systems, two of the discs encircle
single objects that formed shortly before the calculation was stopped,
and the remaining three surround single objects: one star and one brown dwarf,
each of which were ejected during the calculation,
and another substellar object that appears to be in the process of 
being ejected.

Because many of these discs reside in unstable multiple systems 
when the calculation is stopped, we are limited in the conclusions 
we can draw about final disc properties.  However, as discussed 
in Section \ref{BDdisc}, of the 18 brown dwarfs that have been 
ejected and have finished accreting, only one has a resolved disc.
Another object seems to be in the process of being ejected with 
a resolved disc (Table \ref{tablediscs}, object 47)
at the end of the calculation.  Recall from Section \ref{BDdisc} 
that discs around brown dwarfs which are truncated to $\ll 10$
AU in radius would be expected to evolve viscously to a radius 
$\approx 10$ by the time they are observed.  Thus, the vast 
majority of discs should have radii $\approx 10$ AU and only 
a small fraction ($\sim 5-10$ \%) of brown dwarfs should have 
discs significantly larger than this (i.e.\ $>20$ AU).  

Nine stellar
systems (8 single stars and one close binary) have also dispersed from
the cloud.  Only one of the single stars has a resolved disc.  Thus,
these objects also have a low frequency of large discs $\sim 10$\%.  
However, of the remaining 8 discs, 5 surround systems with stellar 
primaries and another surrounds a forming object so isolated 
that it will almost 
certainly become a star.  The two remaining discs surround the 
close binary brown dwarf system (39,41 in Table 2) and a newly formed 
substellar object whose fate is unclear.  Neglecting any further 
evolution of the unstable systems, we can place 
upper limits on the frequencies of discs with radii $>20$ AU
around brown dwarf 
systems of $3/20\approx 15$\% and around stars of $7/15\approx 47$\%.
Thus, given even these highly optimistic estimates, {\it most of the 
stars and brown dwarfs do not retain discs large enough to form our 
solar system.}  

Is this size distribution of discs realistic?  
In a recent study of disc lifetimes in the Taurus star-forming
region, Armitage et al.\ (2002) find that around 30\% of stars lose their
discs within 1 Myr while the remainder have disc lifetimes in the
$1-10$ Myr range.  The latter range is consistent with the lifetimes 
of discs that have an initial dispersion of half an order of
magnitude in mass, but the very short disc lifetimes
require another explanation.  The large fraction of objects 
that suffer severe disc truncation in the calculation presented here
may explain these very short disc lifetimes.

The only star-forming region for which we currently have direct
information on the size distribution of circumstellar discs is the 
Orion Trapezium Cluster.  This is because, unlike most
star forming regions, a reflection nebula behind many of the
young stars enables us to observe discs in silhouette
(O'Dell \& Wen 1994; McCaughrean \& O'Dell 1996; O'Dell 2001).
As discussed in Section \ref{compare}, although our simulation only
forms a small stellar association rather than a cluster
on the scale of the Trapezium Cluster, the resulting 
stellar densities are comparable.  Thus, the types of dynamical
encounters that occur in our simulation may also have occurred
during the formation of the Trapezium Cluster.

We know from infrared excesses that most of the stars and
brown dwarfs in the Trapezium Cluster have discs (Hillenbrand
et al.\ 1998; Lada et al.\ 2000).   Lada et al.\  find
that $\approx 80$\% of the stars in the cluster have excesses 
indicative of discs, while for brown dwarfs the figure is 
$\approx 65$\% \cite{Muenchetal2001}.  Furthermore, 
of $\approx 350$ objects observed by the HST, $\approx 150$
are observed to have discs or proplyds indicating the presence of discs
(O'Dell \& Wen 1994; Stauffer et al.\ 1994; O'Dell \& Wong 1996;
O'Dell 2001).  However, only $\approx 40$ of these exhibit {\it resolved}
silhouette or ionised discs.
HST can resolve discs and proplyds in Orion down to radii
of $\approx 40$ AU.  Thus, although the vast majority of stars
in Orion have discs, the frequency of discs with radii greater
than $\approx 40$ AU is only $\approx 10$\% (Rodmann 2002; 
McCaughrean \& Rodmann, in preparation).  
These numbers are broadly consistent 
with our results.  The implication is that most stars in dense
star-forming environments are incapable of forming large 
planetary systems like our own.

Of course, the Orion Trapezium is a high-mass star forming
region and, as demonstrated by the proplyds, disc destruction
by the O stars may influence the disc sizes along with dynamical
encounters.  Furthermore, the fact that the stellar densities
in Orion are similar to those obtained in the calculation 
presented here does not guarantee that the encounters will be
similar.  Currently in Orion, there is a very low probability of
an encounter at $<100$ AU (Scally \& Clarke 2001).  
For encounters such as those discussed
here to have occured in the Trapezium cluster, the stars must
have formed in small groups which dispersed to form the Trapezium
cluster as it appears today (i.e.\ initially it must have contained
extensive substructure).  Small-scale substructure would have disappeared
by the current age of the cluster (Scally \& Clarke 2002).
Therefore, to test the prediction that most discs are
small due to encounters properly, it is necessary not only to increase 
the resolution
to observe smaller silhouette discs in Orion, but also to
determine the disc size distribution in lower mass star-forming
regions such as $\rho$ Ophiuchus.  This should be possible
with the Atacama Large Millimeter Array (ALMA) or
the Submillimetre Array (SMA).

If the severe truncation of discs that we find is ruled out by future
observations, we will be forced to find ways to reduce the number
of dynamical truncations.  One possibility is that the equation
of state that we use results in discs that fragment too easily.
Not only would this overestimate the number of brown dwarfs
(Section 5.4.1), but it would decrease the number of large discs
for two reasons.  First, some discs that should survive would break 
up into fragments.  Second, the increased number of objects 
would lead to more star-disc encounters and disc truncations.
This possible dependence of the statistical properties on the
equation of state should be investigated in future calculations.

\section{Conclusions}

We have presented the results from one of the most complex 
hydrodynamical star formation calculations to date.  The 
calculation follows the collapse and fragmentation of a 
large-scale turbulent molecular cloud to form a stellar 
cluster consisting of 50 stars and brown dwarfs.  The opacity
limit for fragmentation is mimicked by the use of a non-isothermal
equation of state and the resolution is sufficient to resolve all
fragmentation before this physical limit is reached.  Binary stars 
with separations as small as 1 AU and circumstellar discs with 
radii down to $\approx 10$ AU are resolved.  The calculation 
allows us to study the formation mechanisms of stars and brown 
dwarfs and to determine a wide range of statistical properties
for comparison with observation. 

We find that star formation is a highly dynamic and chaotic process.
A true appreciation of the process can only be obtained from examining
an animation of the calculation.  These can be downloaded from
http://www.astro.ex.ac.uk/people/mbate or 
http://www.ukaff.ac.uk/starcluster.
Fragmentation occurs both in dense molecular cloud cores and in massive
circumstellar discs.  Star-disc encounters form binaries and truncate
discs.  Stellar encounters disrupt bound multiple systems.  The
star formation occurs on the dynamical timescale of the cloud and 
the star formation efficiency across the cloud is variable with a
low global efficiency ($\approx 12$\% when we stop the calculation)
but local efficiencies as high as $\approx 50$\%.  The high local 
efficiencies in dense molecular cores result in bursts of star 
formation because the rapid conversion of gas into stars depletes 
the high-density gas to such an extent that the star formation
essentially comes to a halt until more gas has fallen into the core.  
When enough new gas has accumulated, another burst of star formation occurs.

We find that the opacity limit for fragmentation sets an initial mass
for all fragments of $\approx 0.005$ M$_\odot$.  Subsequently,
the fragments accrete from the surrounding gas.  
Those that manage to accrete enough
mass become stars ($M\gsim 0.075$ M$_\odot$), while the rest are left
as brown dwarfs.  We propose that the initial mass function results
from this accretion process, with each fragment accreting according
to the conditions in which it is formed.  
The calculation produces a mass function that is consistent with 
a Salpeter slope ($\Gamma=-1.35$) above 
0.5 M$_\odot$, a roughly flat distribution ($\Gamma=0$) in the range 
$0.006-0.5$ M$_\odot$, and a sharp cutoff below 
$\approx 0.005$ M$_\odot$.  This is consistent with observational 
surveys.  

Those objects that end up as brown
dwarfs stop accreting before they reach stellar masses
because they are ejected from the dense gas soon after their formation 
by dynamical interactions in unstable multiple systems.  Thus, they
can be viewed as `failed stars'.  This ejection mechanism is very efficient,
producing roughly equal numbers of stars and brown dwarfs
(see also Bate et al.\ 2002a).  However, the close interactions
that occur during these dynamical ejections results in a low
frequencies ($\sim 5$\%) of binary brown dwarf systems. Similarly,
the fraction of brown dwarfs with large (radii $\gsim 20$ AU) 
circumstellar discs is $\sim 5$\%. The accuracy of these frequencies 
is limited by our small number statistics (for example, we can 
only exclude a binary brown dwarf frequency of 20\% at the 
94\% confidence level).  However, further
simulations will increase the significance of the predictions.
Therefore,
observational surveys to determine accurately the frequencies of 
binary brown dwarfs and the sizes of discs around brown dwarfs
should be performed now so that we can test the models.

The calculation produces several binary and higher-order multiple 
systems.  The opacity limit for fragmentation results in an 
initial minimum binary separation of $\approx 10$ AU.  Despite 
this, we find that 7 close binary systems (separations $< 10$ AU) 
exist when the calculation is stopped.  These systems
are produced by the hardening of initially wider multiple systems 
through a combination of dynamical encounters, gas accretion, and/or
the interaction with circumbinary and circumtriple discs (see also
Bate et al.\ 2002b).  These mechanisms lead to close binaries having
a bias towards equal-mass systems and a higher frequency of close
binaries for higher-mass stars.  Many of the close binaries also have
wider companions.  The resulting frequency of close binary systems is
$\approx 16$\%, consistent with observations.  Thus, 
close binary systems need not be formed by fragmentation in situ.

Perhaps the most surprising result of this calculation is that
most of the circumstellar discs in the calculation are severely
truncated by dynamical encounters.  Most young brown dwarfs should 
have discs with radii of $\approx 10$ AU, with large discs 
($\gsim 20$ AU) occurring around only $\sim 5$\% of brown dwarfs.
The discs around many stars are also severely truncated with 
the majority having radii $\lsim 20$ AU (i.e.\ too small to form
our solar system).  Such severe disc truncation, and the associated
low masses, may explain the 
observation that approximately 1/3 of the young stars in Taurus 
have lost their discs when they are only $\approx 1$ Myr old
(Armitage et al.\ 2002).
Currently, the only star-forming region in which
we have information on the size distribution of circumstellar discs
is the Orion Trapezium Cluster thanks to the silhouette discs.  
Our results are consistent with the sizes of discs in the Trapezium
Cluster.  However, massive stars are known to be evaporating 
discs in the Trapezium Cluster and the cluster is much larger than
the system we are able to model.  Thus, we
strongly encourage observations to determine the size 
distribution of discs in low-mass star-forming regions 
such as $\rho$ Ophiuchus.

This is the first of a new generation of star-formation calculations
that resolves all fragmentation and allows us to compare a wide range 
of statistical properties of stars and brown dwarfs with observations.
Future calculations will determine 
the dependence of these properties on the initial Jeans mass in the
cloud, the properties of the turbulence, and will improve the statistical 
significance of the results.
In this way, we hope to understand better the origin of stars and 
brown dwarfs.  

\section*{Acknowledgments}

We thank Mark McCaughrean, Cathie Clarke, and Eduardo Delgado-Donate for 
helpful discussions, and the referee, Ant Whitworth, for useful comments.
The computations reported here were performed using the U.K.
Astrophysical Fluids Facility (UKAFF).

\appendix

\section{Resolving the Jeans mass}

\begin{figure*}
\vspace{-0.5truecm}\centerline{\hspace{1.0truecm}\psfig{figure=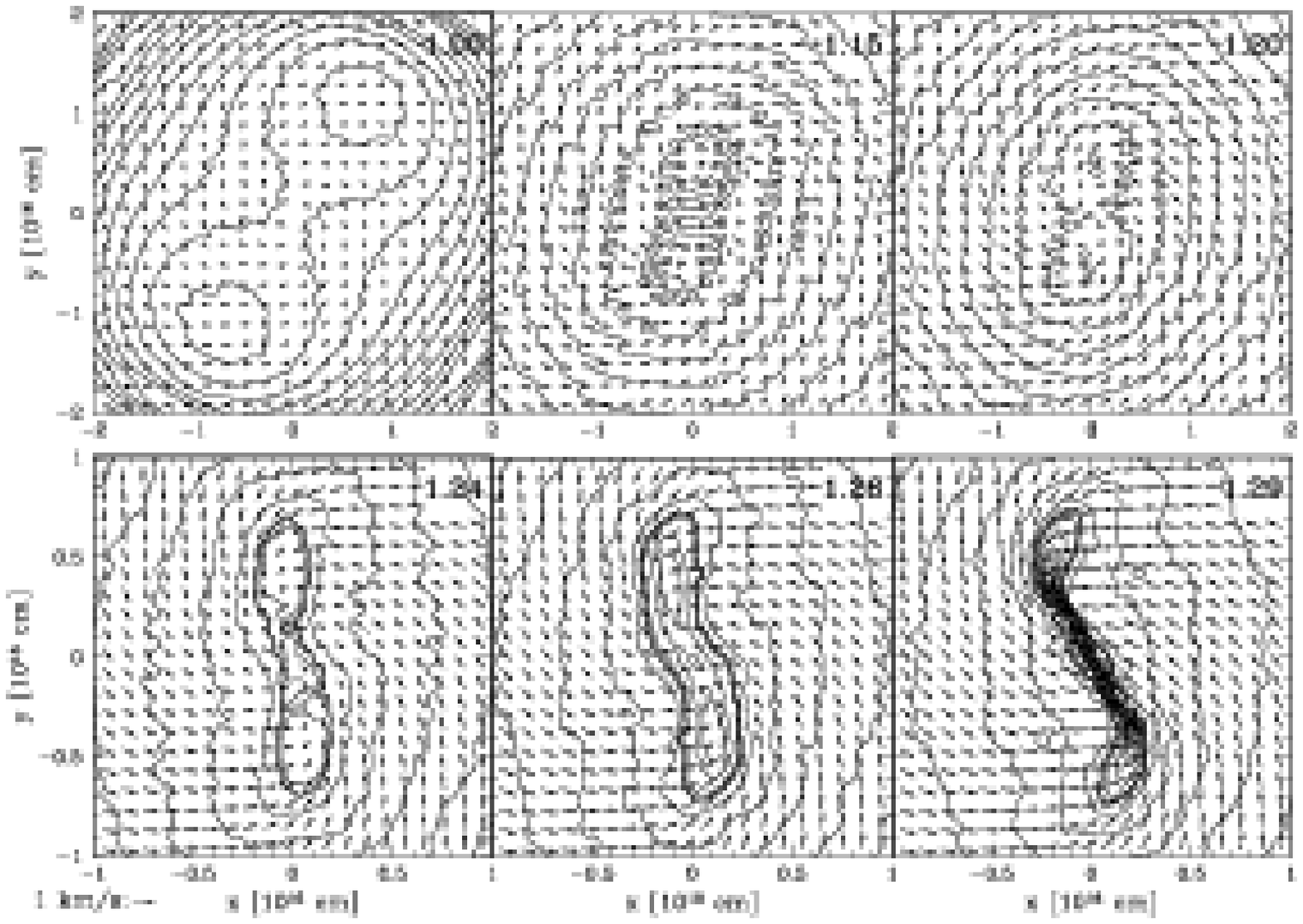,width=17.0truecm,height=17.0truecm,rwidth=17.0truecm,rheight=11.0truecm}}
\caption{\label{appendixfig} Density and velocity in the $x-y$ plane for the binary-bar fragmentation test calculation of Bate \& Burkert (1997) performed with $1.5\times 10^4$ particles.  The figure is constructed in an identical manner to those in Bate \& Burkert (1997) to allow direct comparison.  We find that $1.5\times 10^4$ particles is marginally sufficient to resolve the fragmentation whereas $1.0 \times 10^4$ particles is insufficient.  Density contours are drawn every 1/20 of a decade in the first frame and 1/4 of a decade in the other frames.  In addition, the heavy density contour shows the region within which $\rho > \rho_{\rm res}$.  Velocity vectors are given with length proportional to speed; an arrow representing 1 km/s is given beneath the frames.  Times are given for each frame in units of the initial cloud free-fall time $t_{\rm ff}=1.0774\times 10^{12}$ s.}
\end{figure*}

Bate \& Burkert \shortcite{BatBur1997} performed calculations of
the isothermal collapse of a Jeans-unstable spherical molecular 
cloud core in solid-body rotation with a density distribution
\begin{equation}
\rho = \rho_0 [1+0.1 \cos(m\phi)]
\end{equation}
with $m=2$ and where $\phi$ is the azimuthal angle about the 
rotation ($z$) axis.
The ratios of the thermal and rotational energies to the magnitude
of the gravitational potential energy of the cloud were $\alpha=0.26$ 
and $\beta=0.16$, respectively.
The calculations were performed with a second-order finite difference
hydrodynamics code (Burkert \& Bodenheimer 1993, 1996) and a 
smoothed particle hydrodynamics (SPH) code (Bate et al.\ 1995).

They performed a high-resolution calculation with the grid code
and a series of calculations with different resolutions using the
SPH code.  The initial $m=2$ density perturbation provides two
overdense regions.  As the cloud collapses, these overdense regions
merge into an elongated structure.  The two ends of this elongated
structure each contain more than a Jeans mass so that, as the structure
collapses, it fragments into a binary.  Gas falls into the region 
between the binary, forming a dense gaseous filament.
Thus, the result of the collapse is that the cloud fragments
into a binary separated by a filament of gas.  The subsequent 
evolution of the filament depends on the thermal behaviour
of the gas \cite{BatBur1997}.  

Bate and Burkert obtained good agreement between the grid code 
and the SPH code provided sufficient resolution was used for 
the SPH calculations ($\geq 2\times 10^4$ SPH
particles).  With too few particles ($1\times 10^4$), the collapse
of the elongated structure was delayed and when it finally did 
collapse it formed a filament without fragments at each end.  The 
fragmentation was incorrectly modelled because the Jeans mass at
each end of the elongated structure was not sufficiently resolved.  
Bate and Burkert derived an empirical resolution criterion from 
these calculations that the minimum Jeans mass during a calculation
should be resolved by no fewer than twice the number of neighbouring
particles over which SPH quantities are smoothed 
(i.e.\ $2 N_{\rm neigh}$).

In the calculation presented in this paper, we wish to satistfy this
Jeans mass criterion, but also to model the most massive molecular cloud 
possible.  Bate and Burkert's calculations only pin the resolution
requirement down to between $N_{\rm neigh}$ and $2 N_{\rm neigh}$
particles to resolve the local Jeans mass.  
A factor of two increase in the number of particles results
in an increase of $\approx N^{4/3}=2.5$ in computational time.  
For a calculation that requires $\approx 10^5$ CPU hours, this 
can make the difference between being able to perform the 
calculation and its being impractical.

To obtain a better estimate of the minimum number of particles
required to resolve a Jeans mass sufficiently, we performed an identical
SPH calculation to those presented in Bate \& Burkert \shortcite{BatBur1997},
but using 15000 particles.  The results of this calculation are 
presented in Figure \ref{appendixfig} in an identical manner to the
results in Bate \& Burkert \shortcite{BatBur1997}.  It can be seen
that 15000 particles is also sufficient to resolve the fragmentation
and, thus, the resolution criterion can be further refined to 
$1.5 N_{\rm neigh}$ particles being sufficient to resolve the minimum
Jeans mass.  Following Bate and Burkert, the maximum density for which
the local Jeans mass can be resolved in the test calculation is then given by
\begin{equation}
\rho_{\rm res} \approx \left(\frac{3}{4\pi}\right) \left(\frac{5 R_{\rm g} T}{2 G \mu}\right)^3 \left(\frac{N_{\rm tot}}{1.5 N_{\rm neigh}}\frac{1}{M_{\rm tot}}\right)^2
\end{equation}
where $R_{\rm g}$ is the gas constant, $T$ is the temperature of the gas,
$G$ is the gravitational constant, $\mu$ is the mean molecular weight,
$N_{\rm tot}$ is the total number of particles, and $M_{\rm tot}$
is the total mass of the cloud.  The density contour given by the thick
line in Figure \ref{appendixfig} gives this critical density.
As expected, the fragmentation to form the binary occurs
just as this critical density is surpassed.

Finally, we note that the above test calculation is purely isothermal,
so the Jeans mass decreases monotonically with increasing density.
Thus, the resolution criterion is derived from a calculation
where a Jeans-unstable clump at the resolution limit of the calculation
must continue to collapse and become more and more pronounced.
For the calculation discussed in the main text of this paper, 
the minimum Jeans mass occurs at the density where the equation 
of state becomes non-isothermal (equation \ref{eta}).  
A clump that is marginally Jeans unstable at this density
cannot collapse because, as soon as the gas is compressed, it heats
up and no longer contains a Jeans mass.  Therefore, any clump that
fragments with this equation of state must contain 
more mass than twice the minimum Jeans mass of 0.0011 M$_\odot$
(1.1 M$_{\rm J}$) and, thus, more than $3 N_{\rm neigh}$ SPH particles.

\end{document}